\documentclass[useAMS,usenatbib]{mn2e}
\usepackage{graphicx}
\usepackage{color}
\usepackage{longtable}
\title[A Photometric Catalogue]{A Photometric Catalogue of Quasars and Other Point Sources in the Sloan Digital Sky Survey}
\author[Sheelu Abraham \textit{et al.}]{Sheelu Abraham,$^1$\thanks{sheeluabraham@gmail.com}
Ninan Sajeeth Philip,$^2$\thanks{nspp@iucaa.ernet.in}
Ajit Kembhavi,$^3$\thanks{akk@iucaa.ernet.in}
\newauthor
Yogesh G Wadadekar $^4$\thanks{yogesh@ncra.tifr.res.in}
and Rita Sinha $^5$\thanks{sinharita@gmail.com}\\
$^{1,2}${St. Thomas College, Kozhencheri 689641, India}\\
$^{3}${Inter-University Centre for Astronomy and Astrophysics, Post Bag 4, Ganeshkhind, Pune 411007, India.}\\
$^4${National Centre for Radio Astrophysics,TIFR, Post Bag 3, Ganeshkhind, Pune 411007, India.}\\
$^5${Elviraland 194, 2591 GM The Hague, The Netherlands; formerly with [3]}}

\begin{document}
\maketitle
\begin{abstract}
We present a catalogue of about 6 million unresolved photometric detections in the Sloan Digital Sky Survey Seventh Data Release classifying them into stars, galaxies and  quasars. We use a machine learning classifier trained on a subset of spectroscopically confirmed objects from 14th to 22nd magnitude in the SDSS {\it i}-band. Our catalogue consists of 2,430,625 quasars, 3,544,036 stars and 63,586 unresolved galaxies from 14th to 24th magnitude in the SDSS {\it i}-band. Our algorithm recovers 99.96\% of spectroscopically confirmed quasars  and 99.51\% of stars to i $\sim$21.3 in the colour window that we study. The level of contamination due to data artefacts for objects beyond $i=21.3$ is highly uncertain and all mention of completeness and contamination in the paper are valid only for objects brighter than this magnitude. However, a comparison of the predicted number of quasars with the theoretical number counts shows reasonable agreement.
\end{abstract}


\section{INTRODUCTION}

There has been a surge in the number of large astronomical surveys trying to map the deep sky in terms of its constituents, their number  density and evolution since the early epochs. The Sloan Digital Sky Survey \citep{2000AJ....120.1579Y} is one such survey, covering about a quarter of the sky and providing photometry in five optical bands for $\sim$ 357 million objects in its seventh and final SDSS-II data release \citep{2009ApJS..182..543A}. As less than one percent of these have been spectroscopically observed as a part of the survey, the exact nature of an overwhelming number of objects in the survey remains unconfirmed. This situation will also prevail with future large surveys, where the gap between  imaging and spectroscopy is only expected to widen.  It is, therefore, necessary to develop techniques to identify objects reliably on the basis of their photometric data, using the colours of spectroscopically identified objects as a guide.  In this paper, we will describe the use of a machine learning classifier to classify unresolved objects from DR7 into three categories (quasars, stars and galaxies), using a sample of spectroscopically confirmed objects for training the classifier.

\par The SDSS images objects in five bands with its imaging camera \citep{1998AJ....116.3040G, 2006AJ....131.2332G}, allowing one to view each object in four independent colours. The colours can be used to identify spectral classes of the objects. \citet{2002AJ....123..485S} describe this in detail using SDSS early data release of spectroscopically confirmed objects. The same method is used for the SDSS spectroscopic quasar candidate selection pipeline described by \citet{2002AJ....123.2945R}. This candidate selection procedure, along with followup spectroscopy, has been used to produce a catalogue of over 100,000 spectroscopically confirmed quasars for SDSS DR7, making it the largest available spectroscopic quasar catalogue. SDSS has undergone several improvements in its photometric quality \citep{2008ApJS..175..297A} and other researchers have also made use of colour as a selection tool for preparing catalogues of different objects of interest. These include photometric identification of galaxies \citep{2008ApJ...674..768O}, quasars \citep{2004ApJS..155..257R, 2009ApJS..180...67R}, stars \citep{2007AJ....134.2398C} and photometric redshift \citep{2009ApJ...690...89N, 2009AJ....137.3884R} with reliable accuracies.

\par
\begin{figure}
\includegraphics[scale=0.5]{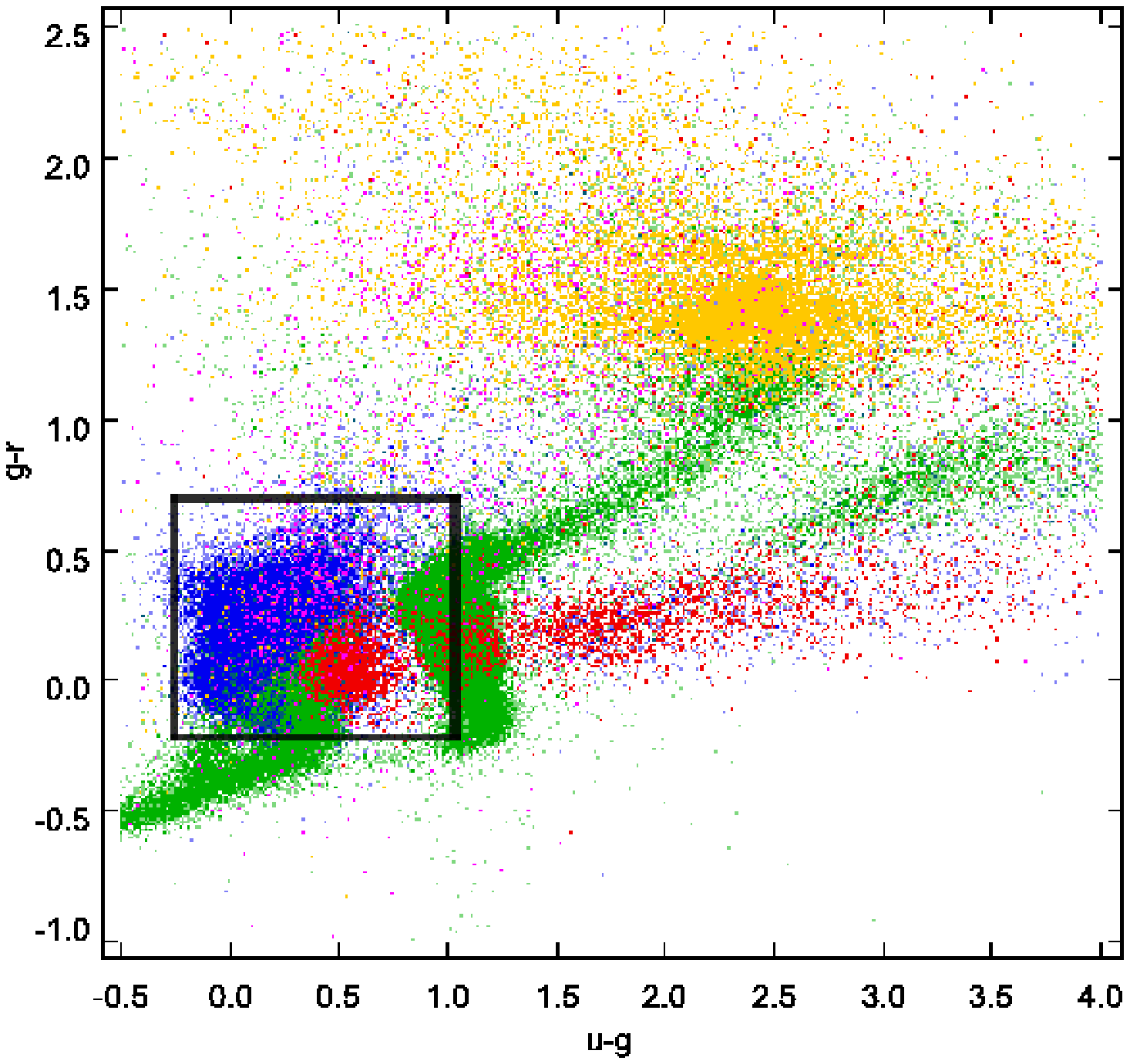}
\includegraphics[scale=0.5]{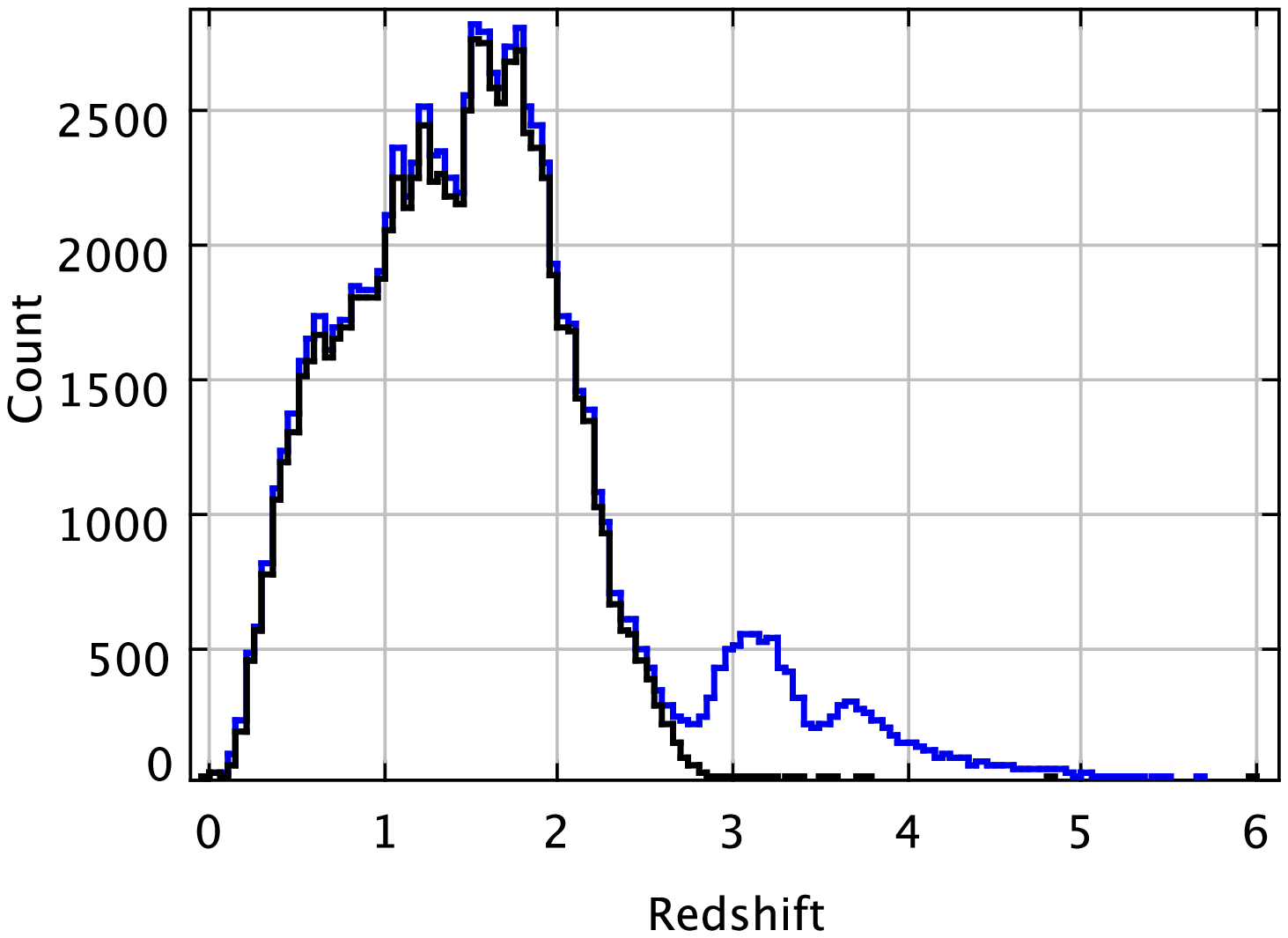}
\caption{Different spectral classes of spectroscopically confirmed unresolved objects in {\it u-g} $-$ {\it g-r} colour-colour space is shown in the upper panel. The black box shows the 2D projection of the region used in the present study. Although it might appear to be dominated by low redshift quasars (blue), this region also has a large number of main sequence stars (green). The SDSS spectroscopic target selection algorithm had efficiently filtered out many stars and that is why they appear in relatively fewer number in the plot. In addition to these, the region is occupied by high redshift quasars (red), a few late type stars (orange) and galaxies (pink). About  84 per cent of all known SDSS and 2dF quasars (blue) are within this colour window (black) covering the full redshift range upto $z \sim$2.6 as shown in the lower panel.}
\label{fig:ug_gr}
\end{figure}

 The colours of quasars and stars are sufficiently distinct for them to be reasonably well separated in a two-colour diagram, say {\it U-B} against {\it B-V}, but there is significant overlap between the two classes and a more reliable separation requires a larger number of colours. Fig.~\ref{fig:ug_gr} shows a typical colour distribution of spectroscopically confirmed ({\it u-g} against {\it g-r}) unresolved objects in the SDSS. It is apparent that the relative number of quasars, stars and galaxies vary widely over different regions of the colour plane. This is a reflection of the fact that objects with intrinsically different spectra occupy different regions of the multi-colour space. However, there is substantial overlap of different types. The black box encloses the region of {\it u-g} and {\it g-r} colour space containing the highest density of the SDSS confirmed quasars. The region has almost equal number of low redshift quasars ($z$ $\le$ 2.3) and stars with a marginal population of late type stars and unresolved galaxies. About 84 per cent of SDSS quasars are within this colour window which also has $\approx$ 46 per cent of all spectroscopic unresolved detections in the SDSS archive. The lower panel of Fig.~\ref{fig:ug_gr} compare the redshift of all known quasars (blue) with the redshift of quasars that fall within the black box (black) to show that this region has a completeness in redshift upto $\sim 2.6$. 

\par  A recent study by \citet{2009AJ....137.3884R} used colours obtained by combining optical data from the SDSS with mid-infrared data to develop an eight dimensional classification algorithm, based on Bayesian kernel density estimation, for the robust selection of quasar candidates.  This method leads to a twenty fold increase in the surface density of quasar candidates, with estimated 97 per cent completeness and 10 per cent contamination, over spectroscopically confirmed quasars in the SDSS. The photometric catalogue (Richards+2009 catalogue) thus produced has identified over a million likely quasars to an order of magnitude deeper than the SDSS spectroscopic limit.

\par We describe a different approach, which uses only SDSS colours and machine learning techniques (see Section \ref{Network}) to provide a large photometric catalogue of unresolved objects, with assigned probability to each object for being a quasar, a star or an unresolved galaxy. This allows candidates of a particular kind with high level of completeness and low contamination to be chosen from the catalogue using an appropriate probability cutoff. We produce a catalogue of objects extending to fainter magnitudes up to SDSS photometric detection limits. For this, we first use a subset of the SDSS spectroscopically confirmed objects covering the region shown as black box in upper panel of Fig.~\ref{fig:ug_gr} to train our machine learning algorithm (here after called the classifier) and then use the trained classifier to test the prediction accuracy on all objects in the same region that have their identity confirmed by the SDSS spectroscopy. We observe that at brighter magnitudes, where the SDSS spectra is available, both stars and quasars can be photometrically separated with a contamination of less than 1 per cent. We also extend our analysis to other surveys where spectroscopic classification is available. Finally we make a prediction on possible class of objects at fainter magnitudes where no spectroscopic confirmation is available. The contamination in those faint magnitudes are uncertain and it will be an acid test for the validity of the proposed method when the newer surveys going to fainter magnitudes will confirm the actual class of those objects. We also briefly discuss how machine learning tools can be used to identify outliers and errors in the data.
\par The organization of the paper is as follows: Section \ref{Data} describes briefly the dataset and the colour selection criteria. The classifier is described in Section \ref{Network}. The construction of training and test data, test results and their analysis are presented in Section \ref{Train}. The catalogue, its format, the results of cross-matching, quasar number density and completeness  are described and presented in Section \ref{catalogue}. Finally, we summarise results in Section \ref{sum}.
\section{THE DATA - SDSS DR7} \label{Data}
The DR 7 catalogue contains spectra of 930,000 galaxies, 120,000 quasars and 460,000 stars \citep{2009ApJS..182..543A}. The SDSS has five filters, namely {\it u, g, r, i} and {\it z} that give spectral coverage from $\sim$ 3900 \AA{} to 9100 \AA{} \citep{1996AJ....111.1748F} with an exposure time of 54.1 seconds per band. Images are taken with a large mosaic CCD camera in drift scan mode \citep{1998AJ....116.3040G}.   We used the 'SpecPhoto' view derived from 'SpecPhotoAll' table of SDSS Catalogue Archive Server (CAS\footnote{http://cas.sdss.org/dr7}) for the training part of this study. 'SpecPhoto' consists of the SDSS spectroscopically confirmed detections with clean spectra. The spectral classification in this table is labelled 'SpecClass' and is given numerical labels ranging from 1 to 6 to represent the different spectroscopic types.

\par The objects used in this study are unresolved point sources which have SDSS {\it i}-band point spread function (psf) magnitude ranging from 14 to 24 and which occupy the colour window region defined by the colour cuts in Table \ref{tab:SDSS1}. This region has 106,466 unresolved spectroscopically classified objects that we  used for training and testing our classifier. This population contains similar numbers of stars and quasars with a small number of faint unresolved galaxies scattered over the region. The data for the photometric catalogue that we describe later also was taken from the same region. This gave us three groups of data. The first two groups, namely, training and testing data have spectroscopic confirmation of their identity. The training data are used to adjust the parameters of our classifier during the training process and the quality of training achieved is assessed using test data. In the testing round, the classifier predicts the  identity of the object based on what it learned from the training data. Since the spectroscopic identities for test data is available, this allows us to determine the completeness and contamination in the predicted classes. All unresolved objects from the region that had spectroscopic confirmation were included in the test data, while a smaller subset of about 10 per cent of it were used as the training data.  The third dataset in the group, referred to as the prediction data was the larger dataset that included all unresolved point sources in the region irrespective of whether or not they had a spectroscopic confirmation. The predictions made on these data are compared with 29 publicly available catalogues to determine the accuracy of our predictions at brighter magnitudes. We describe the details of the training and testing procedure in section \ref{Train}. All magnitudes used in this paper are \"{u}ber-calibrated  psf magnitudes described by \citet{2008ApJ...674.1217P}.
The '\"{u}bercalibration'  improves the photometric fidelity of SDSS data that represent the most robust photometric measurements and was first introduced with DR6 data.  SDSS reports the photometric measurements in asinh magnitudes \citep{1999AJ....118.1406L}. The magnitudes used have all been corrected for galactic extinction.
\section{CLASSIFIER} \label{Network}
We used a Difference Boosting Neural Network (DBNN) \citep{2000cs........6001S} classifier which is a Bayesian supervised learning algorithm. The DBNN has been used in the past for successful star-galaxy classification \citep{2002A&A...385.1119P}, galaxy morphology classification \citep{2002ApJ...568..539O, 2004ASPC..314..617G} and quasar candidate identification \citep{2007HiA....14..609S} problems. Bayes theorem allows one to compute the probability for an event to occur based on some prior belief and a likelihood for the event to be related to an observation. The prior belief is the domain knowledge about the event and usually is the most difficult quantity to  estimate correctly. The estimation of the likelihood also can be difficult when there is conditional dependence between the observations. For example, saying that the colour of an object is red alone does not allow one to say what the object is. Some additional information related by the logical \textbf{AND} operation is to be associated with the colour to make the communication meaningful. We refer to this as conditional dependence. In such situations, the likelihood has to be computed in consideration of all the associated conditions, thereby making Bayesian estimation computation intensive.
\par The DBNN suggests that binning can be used to ascertain conditional independence on the observations. This might appear to be an unrealistic constraint; however, it is not. Classification inherently demands uniqueness in observations. If we bin the observed feature sufficiently narrowly (like a histogram with small bin sizes), each bin would capture the likelihood for the feature to occupy a certain bin location. If we also make separate bins for each class, those will represent the likelihood for the feature to be in a given bin for each class. Now, the value of a feature will always impose some constraint on the possible values the other features can have. The method works if there are sufficiently large number of events (counts) as compared to the number of bins to give a faithful estimate of the likelihood. For this study, our training sample has about 14,356 objects and we use 61 bins for each feature. This is not a critical number and the results would not have changed if we had used a somewhat different number. The procedure is to start with small values and gradually increase the bin size until the different classes and their diversities are adequately captured by the learning algorithm. Since there are no quantitative measures for the adequate capture of diversities, it is often computed, by trial and error, as the best bin size that maximises the prediction accuracy by the classifier. A great advantage of the binning scheme is that conditional independence allows the posterior Bayesian probability to be computed as the product of the individual probabilities and thus significantly simplifies the computational overheads. In addition, the binning scheme allows the classifier to have some of the advantages of non-parametric classifiers while retaining some of the advantages of a parametric classifier that the effective number of features does not grow with the size of the dataset.
\par A second issue that often affects the Bayesian computation is the uncertainty of the prior. The binning scheme has further complicated the estimation of the prior since we now need to know the prior for the likelihood for each bin of the input feature to compute the posterior. DBNN resolves this issue by computing the prior from the data. In the Bayes formula, the prior appears as a multiplicative term. The DBNN initially assumes a flat prior for all the bins. In the training phase, when a set of features are given as input, DBNN makes a prediction about its class based on its current prior and likelihood. If the prediction is wrong, it updates its prior, which is called weight, using a gradient descent algorithm. The important point is that the gradient descent is computed based on the differences in the estimated probabilities for the predicted class and the real class of the sample and hence the computation is largely devoid of fluctuations due to outliers.
\par In principle the classifier is able to compute the probability for the sample to be a member of each of the classes. But in practical situations, the most likely class will have the highest confidence and the second one might have a good share of the remaining confidence. In our software implementation, the classifier is able to make two predictions along with the associated confidence it has in each of the predictions. This information can be extended to identify outliers, reduce contamination and to some extent, evaluate the limitations of the features. 

\par The Bayesian classifier we use can help us to identify unrepresented and rare examples existing in the data. According to the Bayesian theorem, the posterior probability is the normalised product of the likelihood and the prior for an outcome. Likelihood is the probability with which similar events have appeared in the past. Since the likelihood for an unseen event is zero, the classifier will flag it as 'rejected' and will not be classified. Objects with flag 'rejected' can be individually studied and can be subsequently added to the training sample to efficiently identify completely new classes of objects. 

\par The posterior probability will be high when the likelihood and the prior are high. Thus it is often referred to as the 'confidence' in a prediction. A high confidence usually means that the object occupy a location in feature space that is well within the boundary of the cluster formed by its class. However, this is not an assurance that the object is always correctly classified. It may happen that a negligibly small fraction of objects within that cluster belongs to a different class. Because of their small number, the likelihood for them tends to zero and it might happen that such objects will never be correctly identified. On the flip side, this helps the classifier to efficiently learn the boundaries of a class even when there are outliers in the data. Since our classifier bins the data and separately computes the likelihood for each bin, one can optimise the bin width to improve the sensitivity of the likelihood estimates in favour of the marginally represented samples in the data.

The  training and testing procedures for our classifier are explained in the DBNN home page.\footnote{http://www.iucaa.ernet.in/$\sim$nspp/dbnn.html}
\section{TRAINING AND TESTING DATA} \label{Train}
As mentioned earlier, photometric correlations of colours with the spectral class of objects are well established in the literature. The SDSS CAS server has 'SpecPhoto' table that provides the five SDSS magnitudes and spectroscopic classification of all the primary objects selected for spectroscopy by the survey. These spectroscopic classifications are automated in the SDSS pipeline and thus in the case of a small fraction of the objects, the classifications are in error. A follow up visual verification of the classification has thus been carried out for quasars and this is available separately on the SDSS web site as the final SDSS quasar catalogue. In addition to our own visual examination of the spectra, we incorporated the corrections in the SDSS DR7 final quasar catalogue \citep{2010AJ....139.2360S} in our training data.
\par
\begin{figure*}
$\begin{array}{c c}
\includegraphics[width=2.5in]{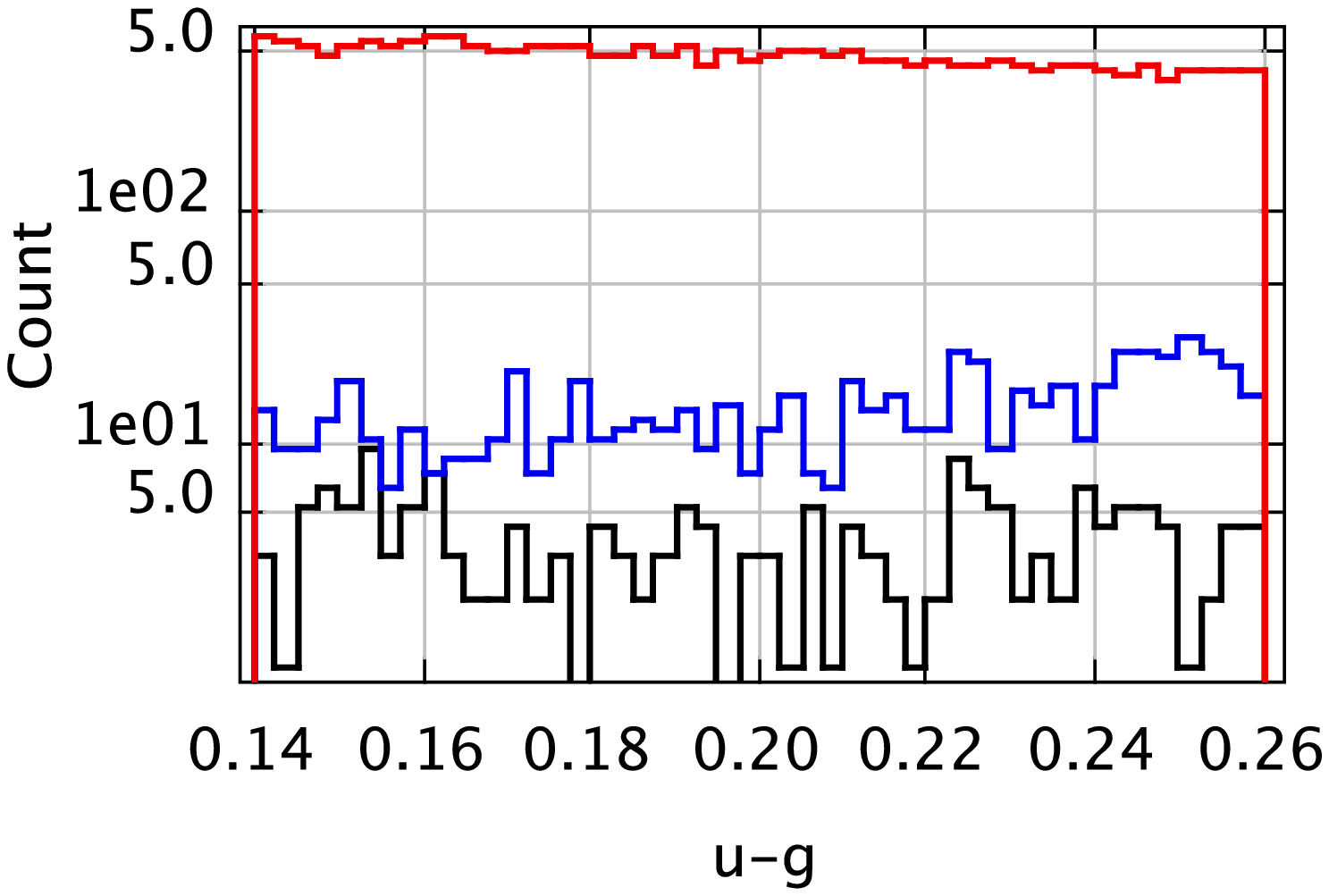} & \includegraphics[width=2.5in]{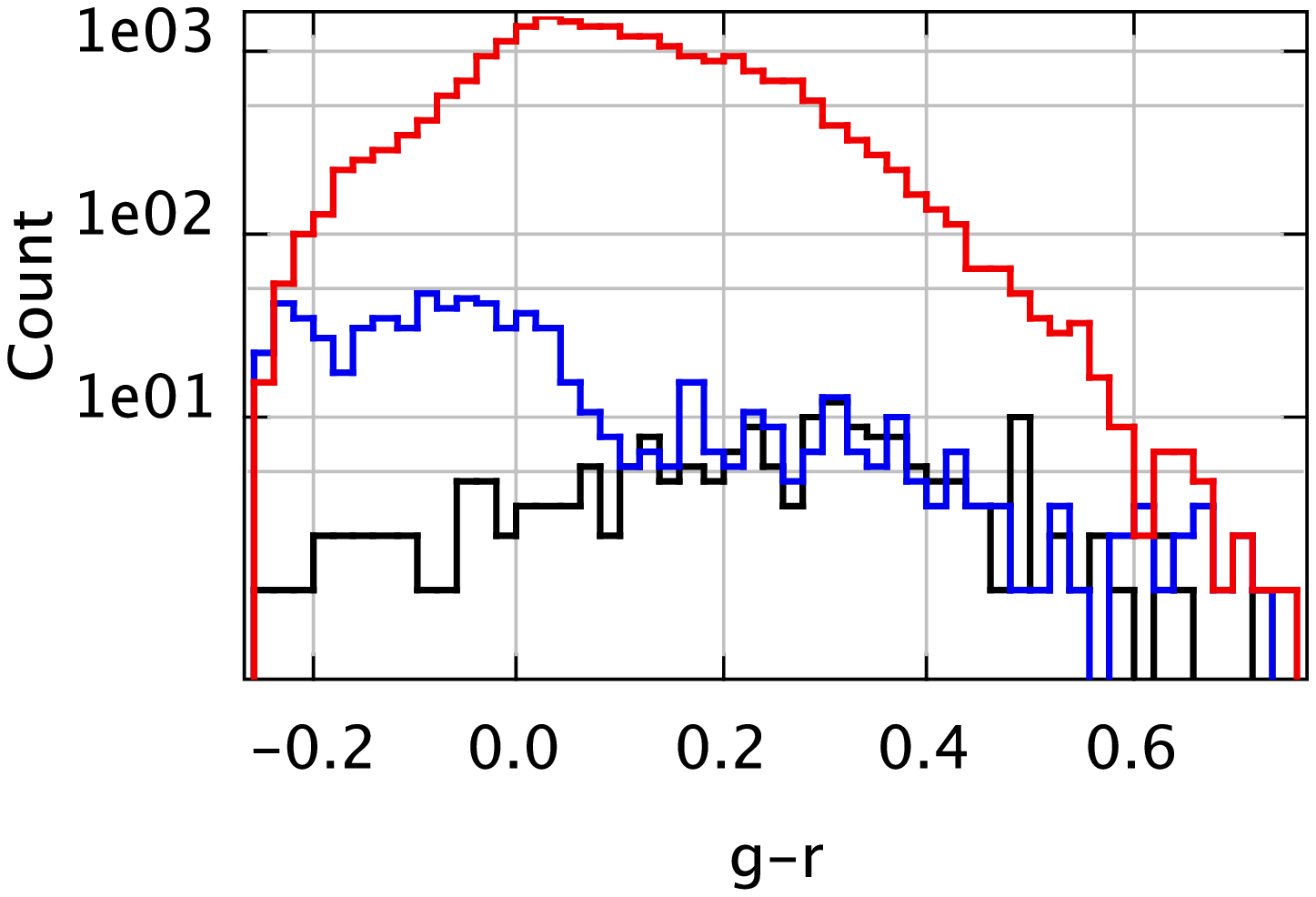} \\
\includegraphics[width=2.5in]{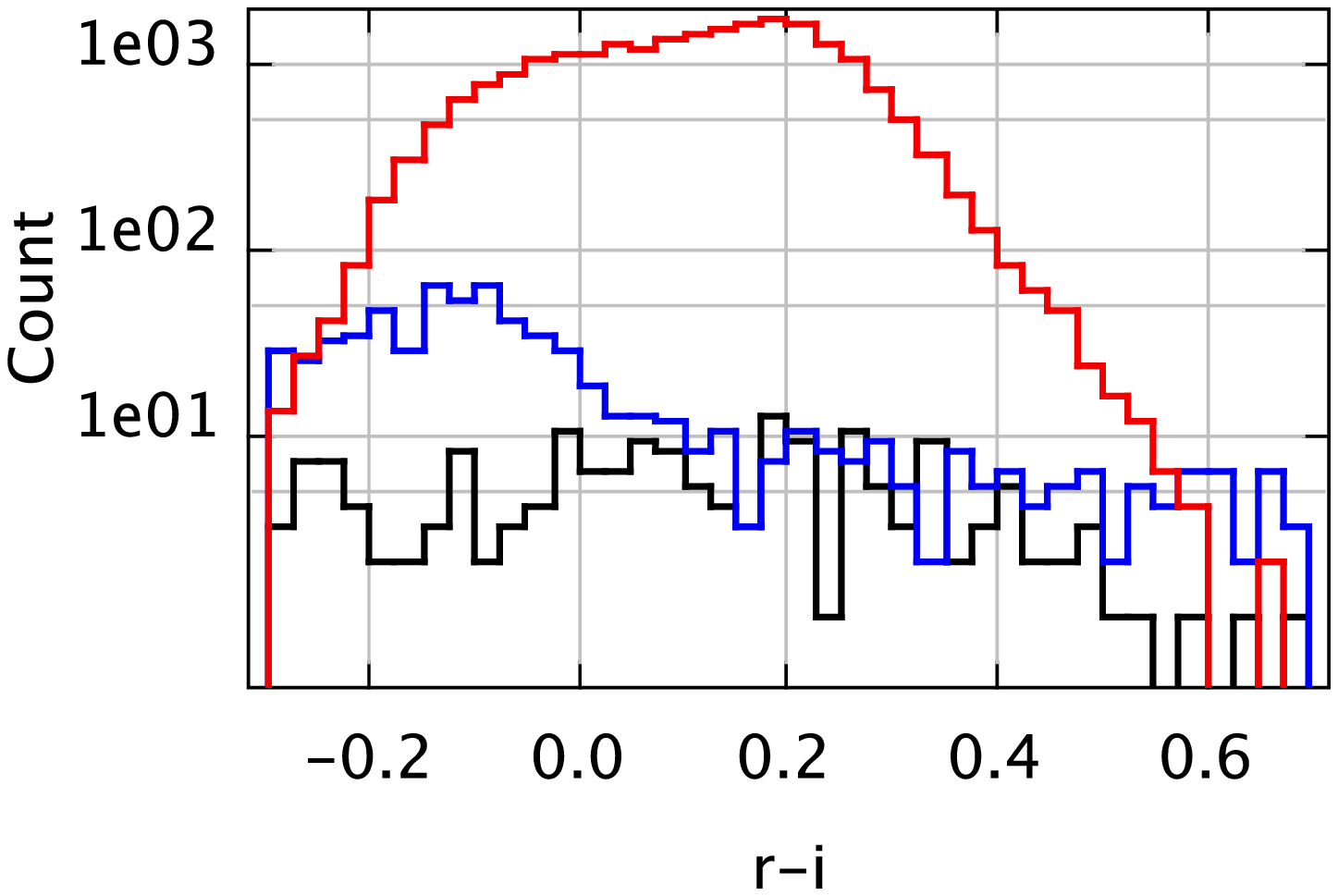} & \includegraphics[width=2.5in]{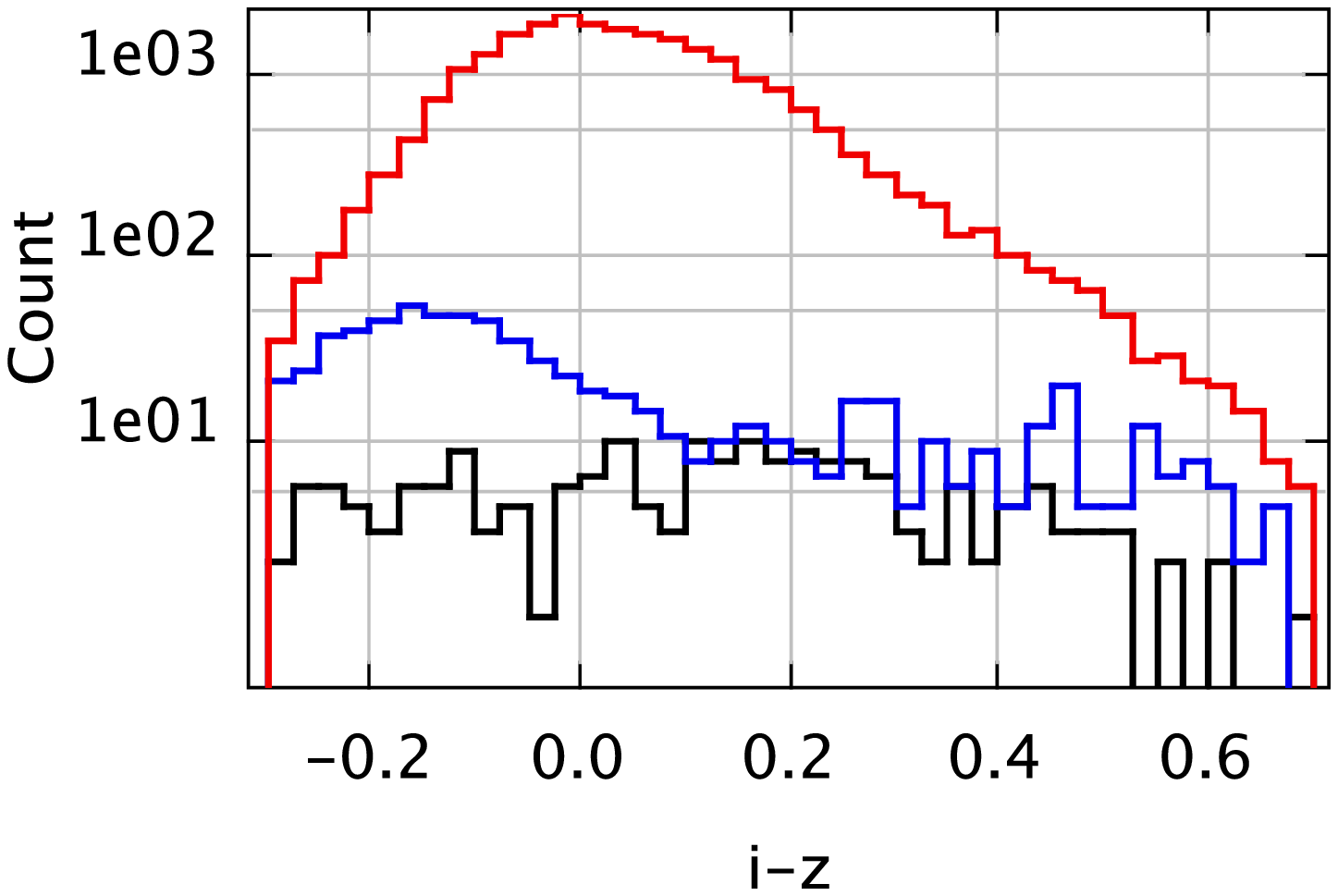} \\
\includegraphics[width=2.5in]{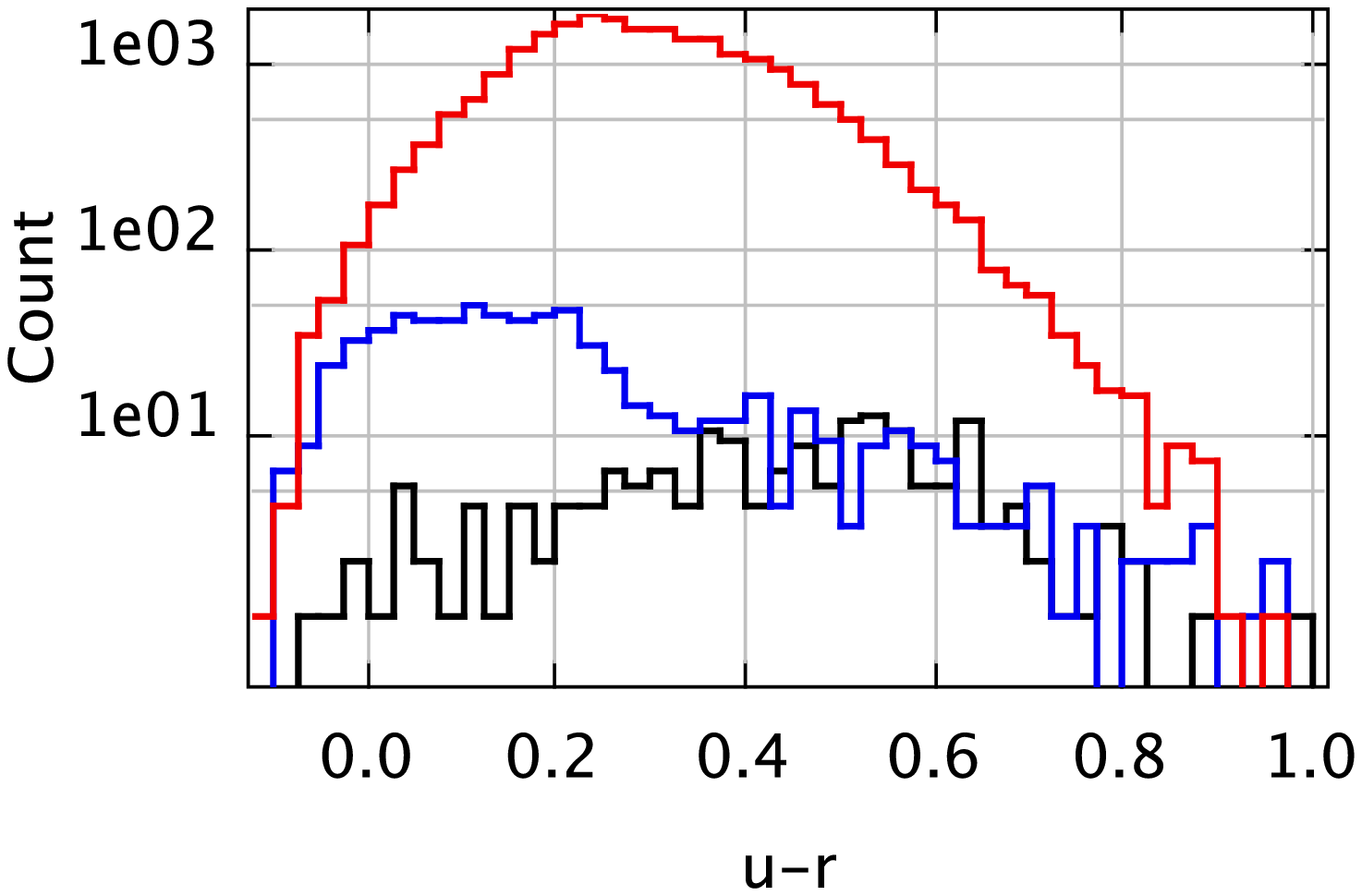} & \includegraphics[width=2.5in]{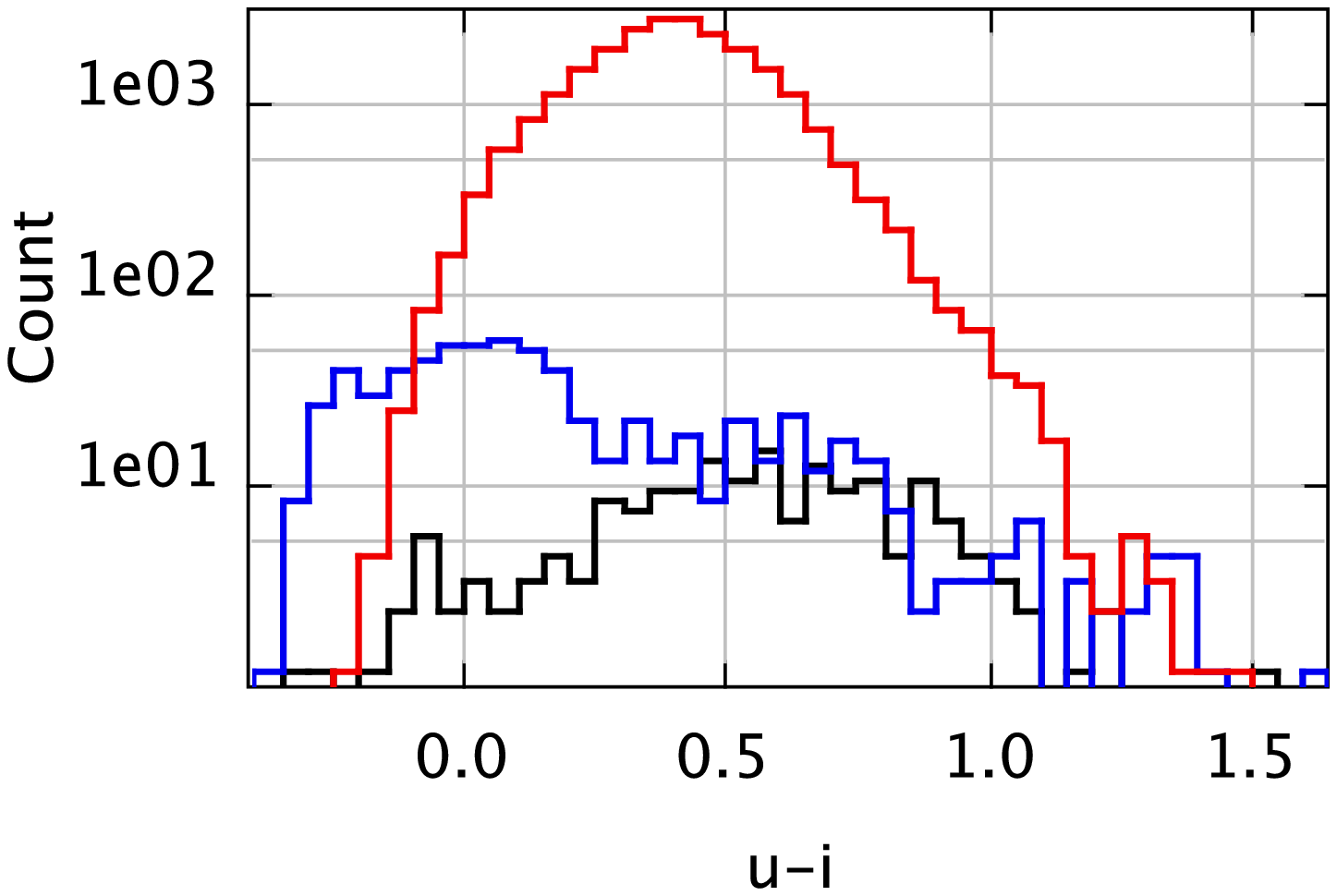} \\
\includegraphics[width=2.5in]{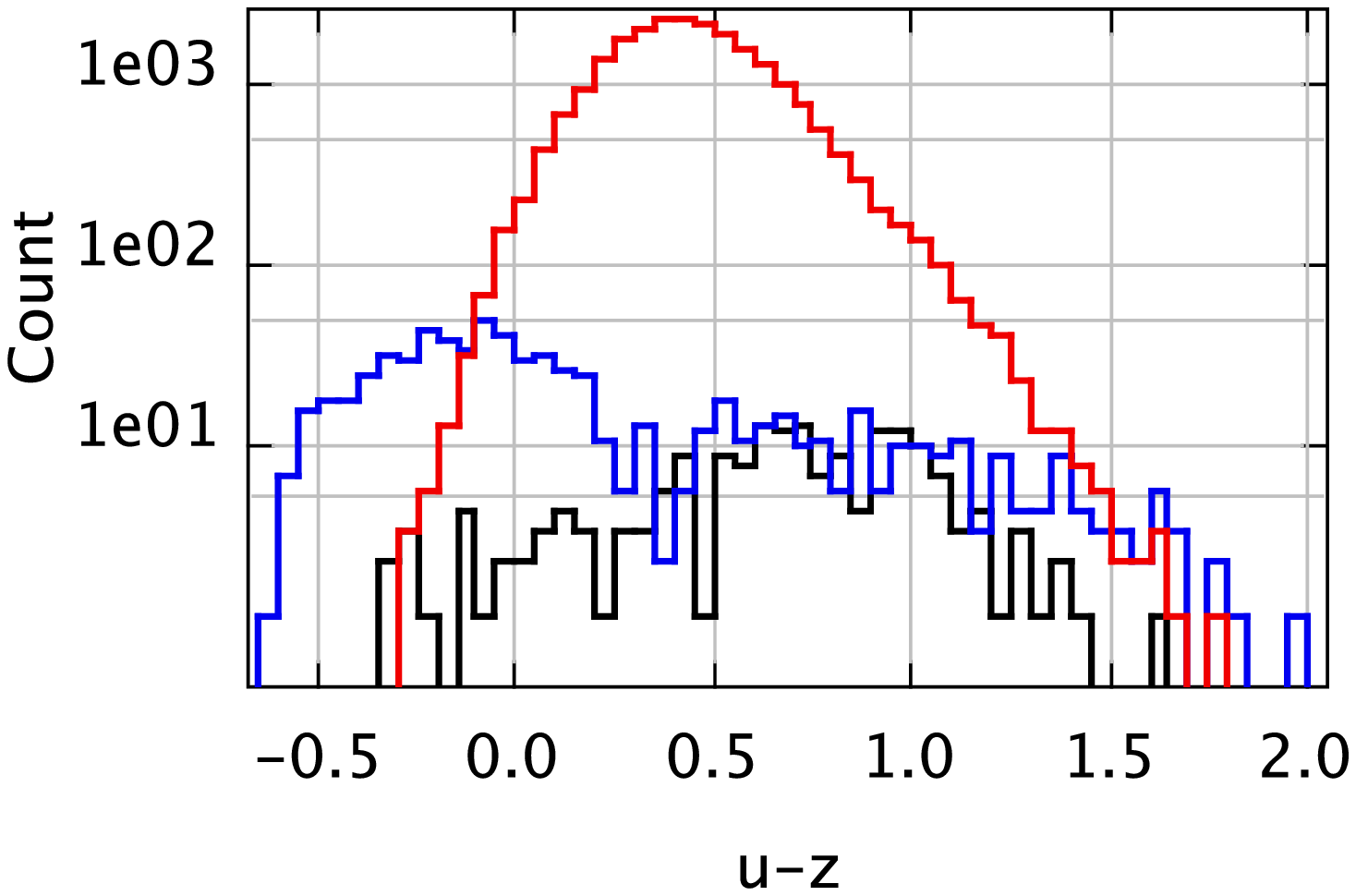} & \includegraphics[width=2.5in]{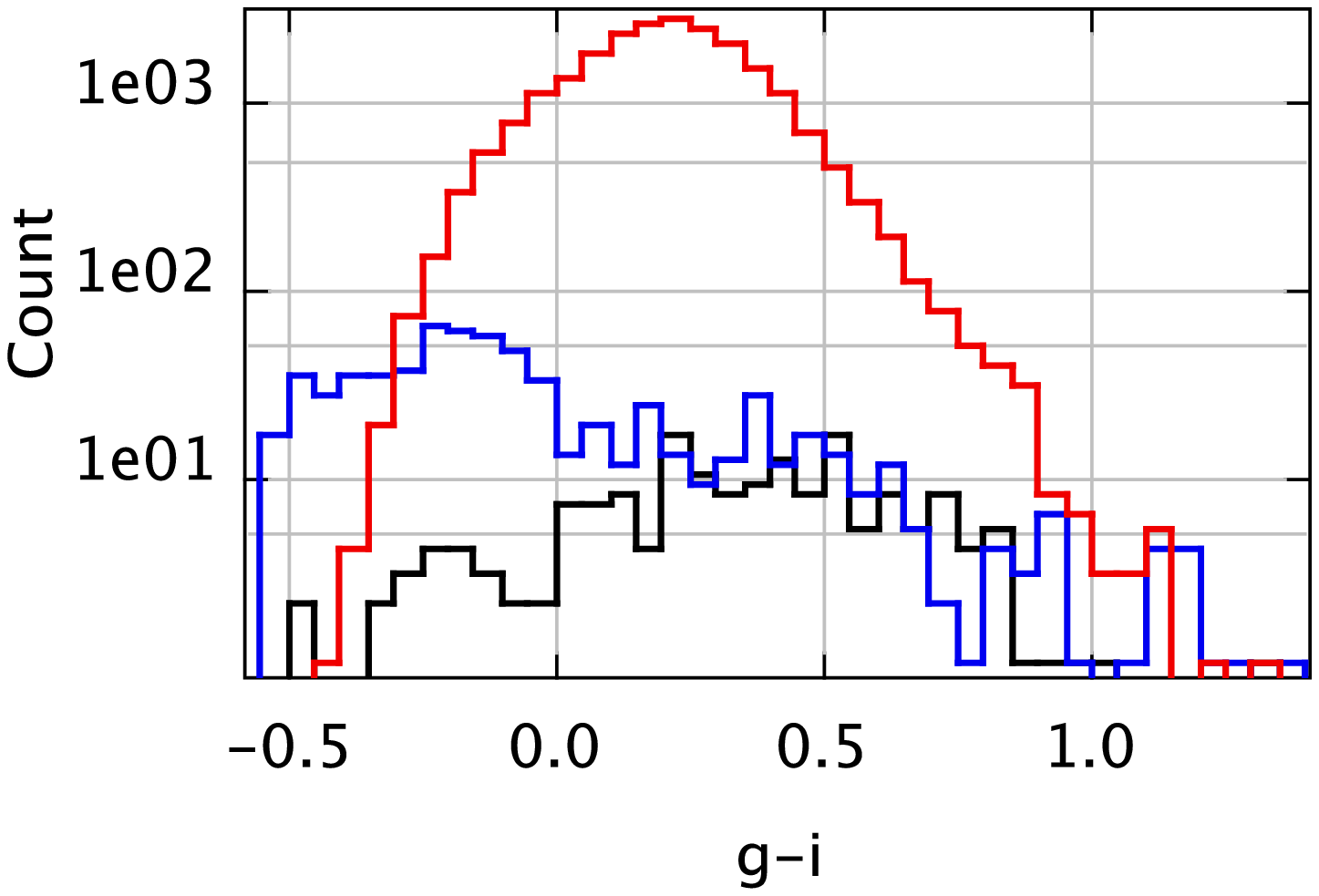} \\
\includegraphics[width=2.5in]{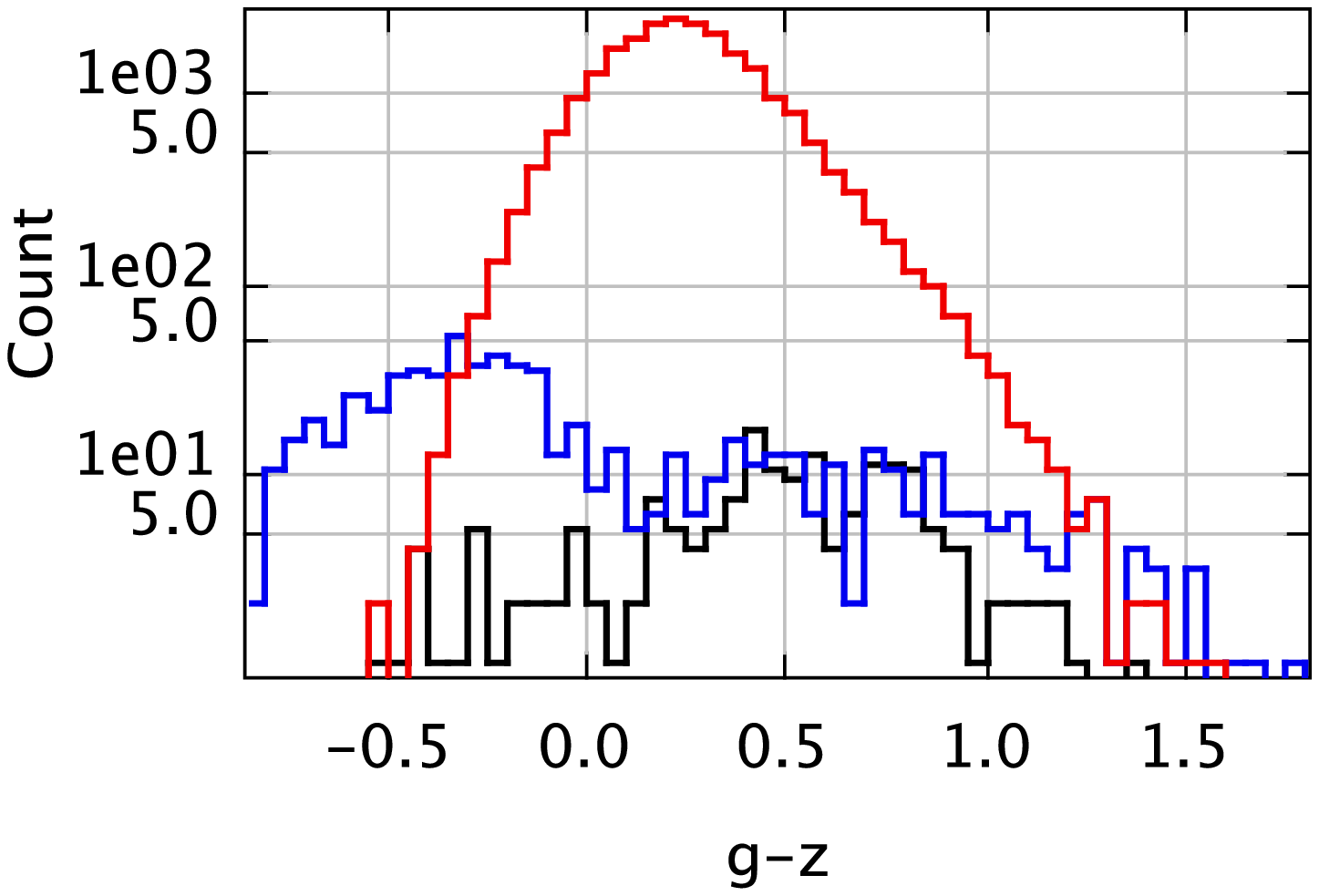} & \includegraphics[width=2.5in]{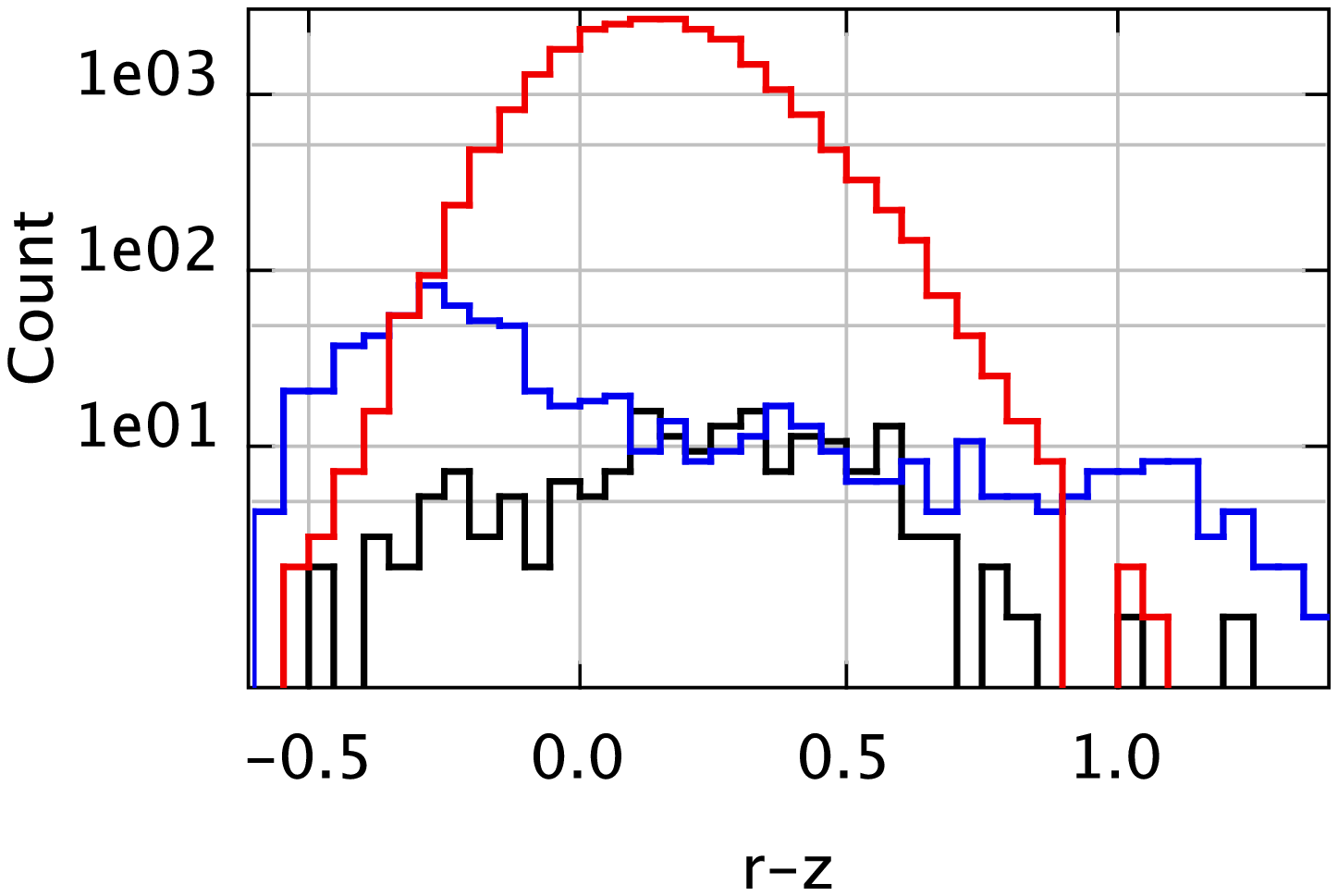} \\
\end{array}$
\caption{Distribution in various colours of spectroscopically identified objects from the region $0.14<${\it u-g}$<0.26$ in the ten dimensional colour space. The colour code is red for quasars, blue for stars and black for galaxies.}\label{fig:10clrs}
\end{figure*}

 The five bands of SDSS can give four independent colours. \citet{2007HiA....14..609S} had shown that good accuracy on quasar classification is possible with the use of the four independent colours and one pivot magnitude with DBNN and our work is an extension of their study. While their classification accuracy on the test data was about 97 per cent, we find that the use of all the ten colours ($^5C_2$) which can be formed from the five SDSS magnitudes, and one magnitude can improve the accuracy to about 99 per cent on the spectroscopically confirmed test data. There is no additional information  in the newly added correlated colours. However, finer details of the probability distribution function (pdf) that are  unresolved when only the independent colours are used become distinct when all the colours are used. A Bayesian classifier differentiates objects based on a likelihood that is estimated from the pdf and hence its resolution plays a significant role in classification. We illustrate this in Fig.~\ref{fig:10clrs} by considering objects from a narrow region, $0.14 <$ {\it u-g} $<0.26$, of one feature and plotting the distributions of the same objects in the remaining nine features. It is seen that despite the fact that some of the colours are correlated, the probability distribution functions looks different in each representation. The Bayesian estimator in our classifier has made use of this, which would be missing if we use only the independent colours, to efficiently separate the overlapping features of objects belonging to different classes.

\par It may be noted that the resolution in the feature space increases when the bin width is decreased. However, narrowing down the bin sizes to improve resolution in colour space requires an unlimited reserve of observations in each bin so that pdfs can be plotted. We thus restrict our analysis to regions in the feature space where maximum spectroscopic classification is available. This is the first criterion we had for selecting the particular window region for our study. Although it is only a small region of the colour space, it has over 46 per cent of the available SDSS spectroscopy in unresolved detections. The colour cuts used by us to define this region are given in Table \ref{tab:SDSS1}.
\begin{table}
\caption{Colour cuts used for preparing training data.\label{tab:SDSS1}}
 \begin{tabular}{lcc}
 \hline
  colour & Lower Limit & Upper Limit\\
\hline
  {\it u-g} &  -0.25 &  1.00 \\
  {\it g-r} &  -0.25 &  0.75 \\
  {\it r-i} &  -0.30 &  0.50 \\
  {\it i-z} &  -0.30 &  0.50 \\
 \hline
 \end{tabular}
\end{table}

\par As said, we use all the ten colours, {\it u-g}, {\it u-r}, {\it u-i}, {\it u-z}, {\it g-r}, {\it g-i}, {\it g-z}, {\it r-i}, {\it r-z}, {\it i-z} plus the {\it u}-band psf magnitude as input features for our classifier. During the training process, by definition, the classifier learns the correlation between colour and spectroscopic types in the training data. This is stored and when similar features are presented to the classifier at a later time, it is used to predict the likely spectroscopic classification of an object. To make comparison easy, the predictions were assigned the same labels as used by 'SpecClass' in the 'SpecPhoto' table of SDSS  (See section \ref{Data}).

\subsection{Data issues and resolutions}
For going to fainter levels, we assume that within the window region of the selected colour space, the likely variations in colour at fainter levels can be learned by the classifier from the colour dispersions observed at brighter magnitudes. However, it is possible that a fainter object is at a different redshift compared to the bright object of its kind and that its observed spectrum and hence colour is altered due to redshift. In the case of quasars, it is known that the redshift distribution at brighter magnitudes ($J <21.2$) are similar to that at fainter magnitudes ($J>21.2$) \citep{1988ApJ...325...92K}. Hence, if we have  spectroscopically confirmed bright quasars from all redshift ranges in our training data, the classifier can learn the intrinsic variations due to redshift differences and then reliably extend this information to classify quasars at fainter magnitudes. This is another reason for restricting our study to a small region in colour space that has maximum spectroscopically confirmed quasars. 
\par The colour cut we used is so selected that it avoids most of the late type stars and faint galaxies that can come in as contaminants in our catalogue. However, to include faint stars and galaxies that might have different colours compared to their brighter counterparts and thus might have entered the colour window, we took a few representative faint objects that have spectroscopic confirmation of their class in 2dF and included them in our training sample. These gave us representative training samples with spectroscopic confirmation to 22nd magnitude in SDSS $i$-band. Our final training data thus had 14,356 unresolved spectroscopically confirmed objects.

For preparing the catalogue, we selected objects that have flag BINNED1 set and excluded objects with flags EDGE, NOPROFILE, PEAKCENTER, SATURATED, NOTCHECKED, PSF\_FLUX\_INTERP, DEBLEND\_NOPEAK, BAD\_COUNTS\_ERROR or INTERP\_CENTER\footnote{http://www.sdss.org/dr7/products/catalogs/flags.html}. At fainter levels, the SDSS magnitude error estimate becomes unreliable. Since the same exposure time is used by SDSS for the entire frame, there will only be a fewer photons from faint objects. This significantly affects the signal to noise ratio and puts an upper limit on the faintness that can reliably be used to extract colour information. Though not compensatory, we restricted the upper limit for magnitude errors to $i \sim$ 0.4  and $g \sim$ 0.2 when preparing the catalogue. However, as SDSS flags are not validated at fainter magnitudes, the level of contamination in our catalogue due to data artefacts is highly uncertain beyond $i\sim 21.3$.

\subsection{Preparation of training data}

The selected region has 106,466 spectroscopically confirmed unresolved objects. For preparing a training sample, we initially took a random set $\sim$ 10 per cent of this data. It was found that the random sampling did not give a good representation of the sparsely represented examples, like late type stars of which there are only 301,  or galaxies that are only about 852 in number in our spectroscopically confirmed data. We used the following strategy to handle this issue. When a feature vector is presented to the classifier during the testing round, it checks whether it has seen an example that looks similar to it earlier. If the test fails, then the classifier will flag that object as 'rejected' without attempting classification. We grouped such flagged objects and added representative samples from them into the training data so that the classifier will be able to classify them. As a result, all the under represented examples got included in the training data.
\par Another issue is caused by redshift that makes the colours of an object appear similar to objects belonging to another class. Since this causes the outcomes for the two classes to be equally likely, the Bayesian probability estimated by the classifier for objects in such regions will be lower. Our classifier use this information to find regions that require extra training samples so that the minor differences in the features may be learned to separate out the classes efficiently. Thus our training data has more examples from regions in the colour space that are occupied by objects from different classes.
\par Another problem we observed in random sample selection was that objects that had higher representation in the data always dominated in the training sample. This causes the dominated class to bias the classifier to its favour. In such a case one has to either remove the excess examples from the training data or add more examples from the under represented class to the training data as a compensatory measure. All these requirements together gave us 14,356 objects with spectroscopic confirmation as our training sample. This is composed of 3,968 stars, 9,236 quasars, 851 galaxies and all the 301 late-type stars. Out of the 14,356 objects in the training data,  2,806 are from 2dF which include 1,025 stars, 1,470 quasars, 278 galaxies and 33 white dwarfs with psf magnitude of {\it i} ranging from 17 to 22 mag. As mentioned earlier, the  2dF objects were added to improve the prediction accuracy of our classifier at the fainter magnitudes where SDSS spectroscopy was not available. Since the training data was constructed from the SDSS and 2dF spectroscopic data, the object type for all of the data are known.

\par During the training process the Bayesian likelihood for each of the training example is computed and related information are stored by the classifier in its runtime file. This information is used later when new data are presented for classification. 

\par To evaluate the performance of the classifier, the trained classifier is used to predict the class of the test data. Since the class of the objects in the test sample is known, the predictions can be easily compared. It is found that the classifier correctly predicted 99.5 per cent of stars and 99.96 per cent of quasars.  In Table \ref{tab:result}, we summarise the actual number of objects in the test data set, the predicted numbers in each class and the accuracy of prediction.
\par In addition to the likely spectral type, the classifier also returns the computed Bayesian posterior estimate for the prediction, which is a measure of the confidence the classifier has in the prediction. Usually an object predicted with high confidence is predicted correctly. But sometimes, it may be noted that an object is  predicted with high confidence to a wrong class. This can happen when the colour of the object becomes similar to the colour of objects in another class which densely populate that region of the colour space. The other possibility is that the assigned object label is incorrect. The latter enabled us to find incorrect spectroscopic labelling of quasars as galaxies in SDSS data, which we describe in the next section.
\begin{table*}
\caption{The accuracy of our classifier as compared to the SDSS DR7 spectroscopic classification of the test sample. \label{tab:result}}
 \begin{tabular}{lllllll}
\hline
Object &\multicolumn{4}{c}{DBNN Predictions}\\
\cline{2-5}
Type & Star & Galaxy & Quasar & Star-Late & Completeness & Contamination \\ \hline
Star & 18,337 & 0 &  90  & 0 & 99.51 \% & 0.47 \%\\
Galaxy & 27 & 705 & 120 & 0 & 82.74  \% & 0.00 \%\\
Quasar & 34 & 0 & 86,852 & 0 & 99.96 \% & 0.28 \%\\
Star-Late & 25 & 0 & 34 & 242 & 80.40 \% & 0.00 \%\\
\hline
Total & 18,423 & 705 & 87,096 & 242 & 99.69 \% & 0.31 \% \\
\end{tabular}
\end{table*}
\subsection{Analysis of Test Failures}
\begin{figure}
{\centering
\includegraphics[angle=270,scale=0.75]{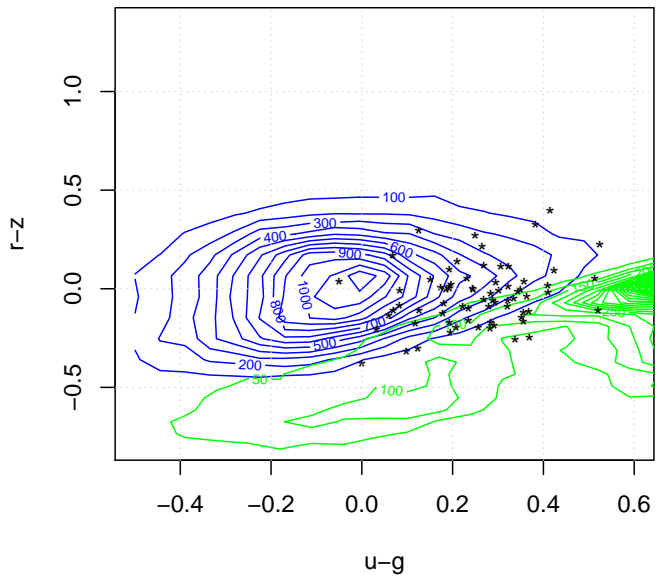}
\caption{A two dimensional projection of the feature space of quasars (blue) and stars (green) along with quasars mistakenly identified as stars and stars mistakenly identified as quasars (black * marks) are shown.The failures are at the colour boundary between quasars and stars in the ten dimensional feature space. }
\label{fig:col}
}
\end{figure}

\begin{figure}
{\centering
\includegraphics[scale=0.5]{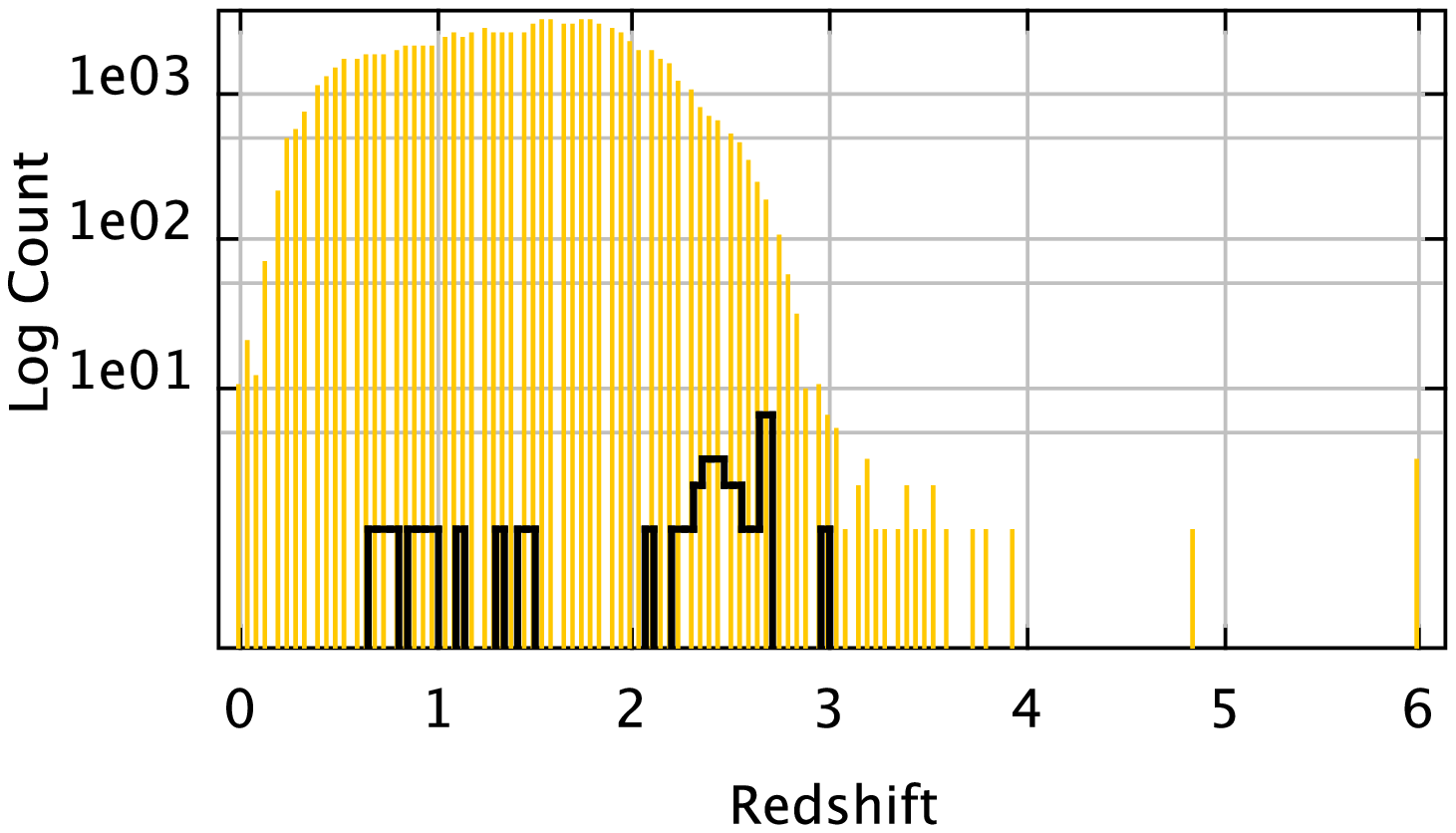}
\includegraphics[scale=0.5]{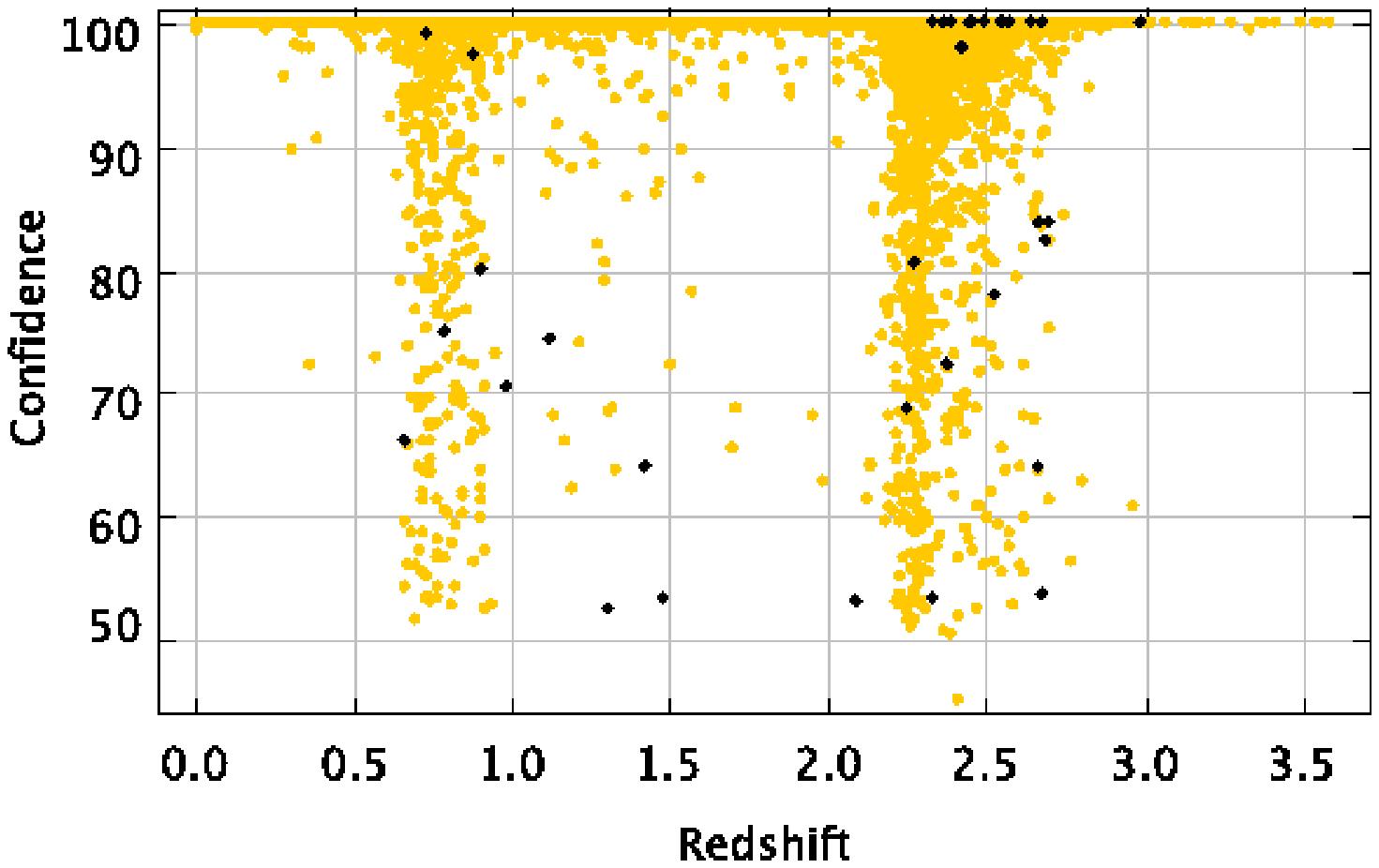}
\caption{The SDSS colours of stars and quasars are indistinguishable in a few patches of redshift. In the upper panel, the histogram in orange indicates the correctly predicted quasars and the histogram in black represents the distribution of quasars that were incorrectly classified as stars (failures). The bottom panel shows how the confidence of the classifier falls in these patches. The black dots represent the confidence of failed objects. It may be noted that, despite the reduced confidence of the classifier in the region shown, most of the quasars were correctly identified. The combined failures from galaxies and stars together is only $\sim$ 0.5 per cent.}\label{fig:dr7Q}
}
\end{figure}

The photometric classification of objects based on colours appears to be straightforward, but it has been observed that colours of different object types sometimes overlap due to various reasons. In our test data there were 34 quasars that got incorrectly classified as stars by our classifier. The locus of the colour feature space that forms the failed cases (black dots) in Fig.~\ref{fig:col} shows that the colours of these objects lie mostly along the boundary of the stars and quasars. For clarity, we did not include galaxies in the plot. What causes this overlap? Fig.~\ref{fig:dr7Q} shows that most of the failures cluster around some specific patches of redshift at which the apparent colours of quasars are similar to those of some dominant stellar populations. Some of these populations, like that near $z$ $\sim$ 0.675, look like white dwarfs \citep{2009ApJS..180...67R}. 
\par In addition to  these, errors in extinction corrections and intrinsic differences between objects within the same spectroscopic class also account for the not so prominent prediction errors. If there are sufficient data, during the process of training, the classifier learns this limitation and assigns a lower confidence to the objects belonging to such overlapping regions in the feature space. This is shown in the lower panel of Fig.~\ref{fig:dr7Q}. By using a confidence cut off, such objects can be removed to obtain a better sample of quasars at the cost of reduced completeness. This is one of the advantages of having a probability estimate attached to each prediction. 
\par The SDSS spectral pipeline had sometimes labelled the objects incorrectly in the SDSS tables. Many of these errors were later corrected in the latest (fifth) SDSS quasar catalogue release \citep{2010AJ....139.2360S}. We initially used the classifications from SDSS tables for training our classifier. On visual examination of the failures, we realised that some of the objects had been incorrectly labelled in SDSS tables and were subsequently corrected in DR7 final quasar catalogue (DR7Q). We thus updated our training sample with the DR7Q classifications and used it to produce the photometric catalogue. However, the corrections in the training data resulted in only very minor changes to our catalogue, less than  0.5 per cent of its previous estimates. This is because changing a few labels in the data need not necessarily bring forth considerable changes in the estimated probability distribution function used by the Bayesian classifier. This advantage of our method helps the classifier to robustly handle unseen outliers that inevitably exist in any data.
\section{THE CATALOGUE} \label{catalogue}
The catalogue is created using our trained classifier on the prediction data that was described earlier. The prediction data are similar to the training and test data with the only difference being that they have no known class label. 
\begin{figure}
{\centering
\includegraphics[scale=0.57]{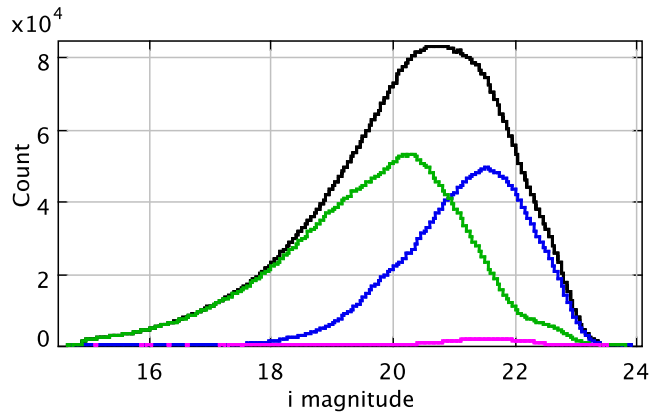}
\includegraphics[scale=0.45]{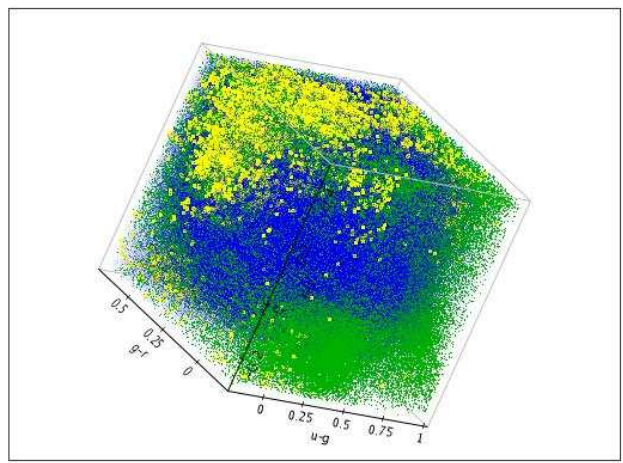}
\caption{Overall magnitude distribution (upper panel) of quasars, stars and galaxies in our catalogue is shown in black extending from 14th magnitude to 24 magnitude in SDSS {\it i}-band. The individual distributions show how the surface density of the types changes. First the stars (green), then quasars (blue) and galaxies (pink) peak in the distributions as we move towards fainter magnitudes. In the lower panel, a 3D colour cube [u-g, g-r, r-i] of the 6 million predictions in our catalogue, colour coded as galaxies (larger yellow points are used to make their locus visible), stars (green) and quasars (blue), is shown.}
\label{fig:ihist}
}
\end{figure}

\begin{figure}
{\centering
\includegraphics[scale=0.57]{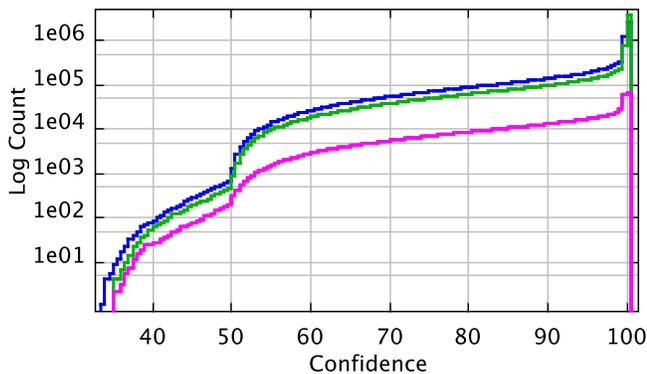}
\caption{Plot of the normalised cumulative histogram of the predicted Bayesian posterior probability for quasars (blue), stars (green), galaxies (pink) in the catalogue are shown. The individual values with each prediction can be regarded as the confidence the classifier has in that prediction. This information may be used to subgroup objects for follow up studies on the basis of similarities.}
\label{fig:confhist}
}
\end{figure}
\par There are 6,038,247 rows (object detections) in our catalogue. These are classified into 2,430,625 quasars, 3,544,036 stars and 63,586 galaxies by our classifier.  The distribution of {\it i} magnitude of the objects in our catalogue is shown in Fig.~\ref{fig:ihist}. According to the classifier predictions, the distribution of stars peaks at {\it i} $\sim$ 20, followed by quasars and galaxies at $\sim$ 21.5 with truncation at fainter magnitudes due to the photometric limit of the SDSS survey in our colour box. It may also be noted that the colour distribution of the 6 million predictions in the catalogue looks similar to that shown in Fig.~\ref{fig:ug_gr}. The catalogue contains 3,412,751 (57 per cent) objects fainter than psf {\it i} magnitude of 20.2, which is the limiting magnitude of SDSS quasar spectroscopy \citep{2002AJ....123.2945R} and of these 1,99,690 are predicted as quasars, 1,352,871 as stars and 60,190 as faint galaxies.
\begin{table}
\caption{The distribution of objects in our photometric catalogue. \label{tab:cat}}
 \begin{tabular}{ll}
\hline
Class & Predicted No\\
\hline
Main Sequence Stars & 3,540,337 \\
Galaxy & 63,586 \\
Low Z Quasars & 2,257,905 \\
High Z Quasars & 172,720 \\
Late type Stars & 3,699 \\
\hline
Total & 6,038,247 \\
\end{tabular}
\end{table}

\par It was found that 69 per cent of the objects in the catalogue are predicted with 100 per cent confidence while only 4 per cent  have less than 90 per cent and the remaining 27 per cent have confidence between 90 and 100 per cent. This means, approximately 69 per cent of the objects are well resolved in the colour space while the remaining 31 per cent where sharing the colour space with objects from different classes. Only 1 per cent objects were predicted with confidence less than 70 per cent, which are from regions where there is substantive overlap between different classes. As stated earlier, lower posterior prediction probability occurs when the events are equally likely, say, the colour of one type of object overlaps with another in the data. A plot of normalised cumulative histogram of these probabilities in the catalogue is shown in Fig.~\ref{fig:confhist}. 

\begin{table*}
\caption{Sample rows from our photometric catalogue based on SDSS DR7 (Please see text for column references). \label{tab:samptab} }
 \begin{tabular}{cccccccc}
\hline
SDSS ID & R.A. & DEC.  & i mag & Most Probable & Confidence 1 & Second Most Probable& Confidence 2 \\
 &  &  & &  &  \% &  &  \% \\
\hline
587732772667326484 & 185.24782587 & 10.87881822 & 18.59683 & 3 & 100.0000 & 1 & 0.00000 \\
587732772667326517 & 185.18802403 & 10.93259451 & 19.817682& 1 & 99.99999 & 4 & 0.00001 \\
587732771042623616 & 152.67642584 & 8.17636451 & 19.76019 & 2 & 100.0000 & 3 & 0.00000 \\
587732771039346749 & 145.17463514 & 7.29821365 & 18.57126 & 3 & 100.0000 & 4 & 0.00000 \\
587732771039871127 & 146.37899605 & 7.54401918 & 19.97955 & 4 & 98.93777 & 3 & 1.06223 \\
587732771043016884 & 153.62477833 & 8.34608604 & 19.97215 & 1 & 100.0000 & 3 & 0.00000 \\
587732771042296010 & 151.99741353 & 8.11097072 & 19.46754 & 2 & 88.48774 & 3 & 11.51226 \\
587732771039608895 & 145.81816028 & 7.50764781 & 18.41648 & 3 & 100.0000 & 1 & 0.00000 \\
587732771047866444 & 164.72707514 & 9.09094786 & 19.79277 & 4 & 99.99086 & 3 & 0.00914 \\
587732771043410105 & 154.58767573 & 8.27519667 & 19.55149 & 3 & 95.53498 & 1 & 4.46502 \\
\hline
\end{tabular}
\end{table*}

\par Objects with the same confidence value within a small hypercube of the feature space forms the most similar objects in the entire data. One can build subset of objects grouped on the basis of their confidence value for follow up studies. For example, to study objects of a specific kind, in addition to other regular identifiers like colour etc, the confidence measure given in our catalogue for its kind can be used as an additional dimension.

\subsection{Catalogue Format}
The full catalogue contains 8 columns of 6,038,247 SDSS sources that uses the same numeric labels in SDSS tables to  classify objects into stars, galaxies, low redshift quasars, high redshift quasars (HizQSO) and late-type-stars. The quasars in the catalogue include BAL quasars, BL Lacertae objects and other AGNs. Although we assign separate labels for high and low redshift quasars, in this first release, we are not classifying the different types of AGNs and its sub classes and are grouping them all under the single name quasars. A summary of the objects predicted by the classifier is given in Table \ref{tab:cat}. A sample of  10 entries in the catalogue is given in Table \ref{tab:samptab}. The content in each column is as follows.
\begin{enumerate}
 \item SDSS ID : SDSS photometric object ID
\item R.A. : Right ascension in decimal degrees (J2000)
\item DEC. : Declination in decimal degrees (J2000)
\item i mag : SDSS $i$ -band PSF \"{u}bercalibrated asinh magnitude
\item Most Probable : Most probable class of the object, represented by integers with the same meaning assigned by SDSS in their tables.
\item Confidence: The confidence the classifier has in the most probable class.
\item Second Most Probable : The second most probable class of the object, same format as Object Type.
\item Confidence: The confidence the classifier has in the second most probable class.
\end{enumerate}

\begin{table*}
\caption{Summary of the matching of our catalogue predictions with some existing catalogues.\label{tab:cross1}}
\begin{tabular}{ccccccccccc}
\hline
&  \multicolumn{3}{c}{DBNN Predictions} &  & \multicolumn{3}{c}{Failures as per catalogue} \\
\cline{2-4} \cline{6-8} \\
Cat. Code           & Quasar       &
Galaxy     & Star  &
&     Quasar     &
Galaxy     & Star     &
Accuracy & {\it i} mag Range & Ref\\
\hline
2DF &5976 &238 &1535 & &122 &0 &52 & 98\% & 17.0 - 22.0 & 1 \\
XBH & 212 & 0 & 0 & &0 &0 &0 & 100\%  &15.8 - 20.5 & 2 \\
ASFS &1088 &12 &31 & &0 &12 &31 & 96\% & 14.5 - 22.1 & 3 \\
BATCS& 21 & 0 & 0 & &3 &0 &0& 86\% & 18.1 - 20.5 &4 \\
CGRBS &265 &1 &0 & & 0 & 1 & 0 &100\% & 14.7 - 21.5 & 5 \\
DLyaQ& 21 &0 & 1 & &0 &0 & 1 & 95\% &16.5 - 19.4 & 6 \\
F2QZ &186 &1 &3 & &0&1 &3 &98\% & 16.6 - 21.0 & 7 \\
KFQS &144 &2 &13 & &3 &1 &7 &94\% & 16.8 - 20.6 & 8 \\
LQAC &61504 &17 &267 & &0 &17 &267 &100\% & 14.7 - 22.3 & 9 \\
LQRF &60280 &14 &219 & &0 &14 &219 &100\% &14.7 - 21.7 &10 \\
BZC &249 &4 &2 & &0 &4 &2 &98\% & 15.0 - 21.0 & 11 \\
PCS &53 &0 &2 & &0 &0 &2 &96\% & 15.1 - 18.5 & 12 \\
ROSA &1134 &0 &1 & &0 &0 &1 &100\% & 15.5 - 20.5 & 13 \\
SQ13 &65223 &55 &395 & &0 &55 &395 &99\% & 14.7 - 22.8 &14 \\
SQR13 &7 &0 &21 & &7 &0 &0 &75\% & 16.3 - 20.3 &14 \\
DR7Q &79140 &17 &341 & &0 &17 &341 &100\% &14.9 - 21.8&15 \\
SSSC &82 &2 &1171 & &82 &2 &0 &93\% & 14.9 - 21.5 &16 \\
SSA13 &5 &0 &1 & &0 &0 &0 &83\% & 17.8 - 20.8 &17 \\
XMMSS &37 &0 &5 & &1 &0 &2 &93\% & 14.9 - 20.7 &18  \\
SDSS/XMM & 580 &0 &0 & &0 &0 &0 &100\% & 15.2 - 20.5 &19  \\
RASS/2MASS &6 &0 &0 & &0 &0 &0 &100\% & 15.5 - 18.4 & 20 \\
CAIXA & 16 &0 &0 & &0 &0 &0 & 100\% & 15.1 - 17.8 & 21  \\
WDMB &20 & 0 & 106 & &20 &0 & 0 & 84\% & 15.3 - 20.5 &22 \\
PMS &639 &6 &19596 &  &639 &6 & 0 & 97\% &14.8 - 20.2 & 23 \\
GLQ &2 &0 &0 & &0 &0 &0 &100\% &18.8 - 19.1 & 24 \\
\hline
\end{tabular}
\\(1) Croom et al.2009a; (2) Kelly et al.2008; (3) Healey et al.2007;
(4) Zhang et al.2004; (5) Healey et al.2008; (6) Curran et al.2002; (7) Cirasuolo et al.2005; (8) Maddox et al.2008; (9) Souchay et al.2009; (10) Andrei et al.2009; (11) Massaro et al.2009; (12) Kuraszkiewicz et al.2004; (13) Suchkov et al.2006; (14) Veron-Cetty \& Veron 2010; (15) Schneider et al.2010; (16) Skiff 2009; (17) Fomalont et al.2006; (18) Watson et al.2009; (19) Young et al.2009; (20) Haakonsen \& Rutledge 2009; (21) Bianchi et al.2009; (22) Heller et al.2009; (23) Gould \& Kollmeier 2004; (24) Oguri et al. 2008;

\end{table*}
\subsection{Cross-Matching}
We carry out cross matching between our catalogue and several other catalogues which contain a subset of objects from our catalogue. The purose of the cross matching is two fold:
\begin{enumerate}
\item To identify potential limitations of the catalogue based on available data. This is important because the whole exercise is based on a single survey and has not considered any of the inherent constraints in the survey. In many surveys the target selection algorithms are optimised to detect a particular category of candidates and when the training data is derived from it, it is not necessary that it would be representative for the kind of objects observed by other surveys. Cross matching can reveal such biases if they exist.
\item To estimate the quality of classification at brighter magnitudes where existing spectroscopy can provide a representative sample. This is significant because, even with modern technology, spectroscopic confirmation of all bright objects is impossible and one has to adopt other methods to determine their number density of various types. Since different surveys cover a different set of objects depending on their objective,  robustness of a method can be estimated by cross matching predictions with spectroscopic confirmations done by different surveys.
\end{enumerate}

\par To this end, we match our predictions with several other catalogues which contains some of the objects in our catalogue and summarise the results in Table \ref{tab:cross1}. The tables also include the magnitudes covered by the matched objects in the catalogues. The list of unconfirmed predictions are given in Table \ref{tab:cross2} and a detailed discussion of the cross validation results are given in Table \ref{tab:discript}. Multi-wavelength catalogues covering X-ray, optical and radio with spectroscopically confirmed objects were given preference in selecting catalogues for cross matching.

\par All cross-matching is done by matching the celestial coordinates of the objects within 1 arcsec of their value in our catalogue. Since the selected region is rich in quasars, we selected a few spectroscopic surveys with confirmed quasars as reference data for our catalogue. The surveys that we used for this are marked with a ** in Table \ref{tab:discript}. This resulted in the identification of 90,249 spectroscopically confirmed quasars with objects in our catalogue. We find that we had correctly classified 89,549 ($\sim$ 99 per cent) of the objects as quasars. These included 10,230 ($\sim$ 11 per cent) spectroscopically confirmed non-SDSS quasars, of which, 9,887 (97 per cent) were correctly classified while labelling 48 of the objects as galaxies and 295 as stars. In the 9,887 non-SDSS quasars, 5,167 were fainter than SDSS quasar spectroscopic upper limit of 20.2 and 5,036 (97 per cent) were correctly predicted as quasars by our classifier. Comparing our catalogue with spectroscopically identified stars from 2dF, Skiffs catalogue of Stellar Spectral Classification, Proper Motion Catalogue from SDSS $\cap$ USNO-B (details and references in Table \ref{tab:discript}) and SDSS DR7 catalogue resulted in  36,645 stars in the window region selected by us. Of these 35,830 ($\sim$ 98 per cent) stars were correctly identified by our classifier. Comparison with 746 spectroscopically confirmed galaxies from SDSS DR7 and 2dF resulted in the correct identification of 546 galaxies.
\par Comparison with DR7Q gave 79,498 quasars of which 79,140 were correctly identified by our classifier. The 341 quasars that were predicted as stars by our classifier are from a few patches of redshift that includes $z$ $ \sim$ 0.675 and 2.3 \citep{2009ApJS..180...67R, 2007AJ....134..102S}, where the colours of quasars are known to merge with the colours of stars. These regions are identical to what was shown in the upper panel of Fig.~\ref{fig:dr7Q} where, orange colour represent correctly predicted quasars and black represent quasars incorrectly classified as stars. 

We found that our classifier could correctly recover difficult and interesting objects, like  75 per cent of the rejected objects in the  13th edition of the Quasar and Active Galactic Nuclei catalogue \citep{2010yCat.7258....0V}. These objects were earlier counted as quasars and were recently identified as belonging to some other class. 

\par
\begin{figure}
\includegraphics[scale=0.5]{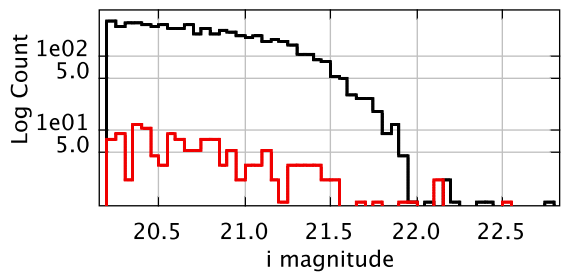}
\caption{The histogram of quasars in our catalogue that are fainter than SDSS spectroscopic magnitude limit in i-band and are having spectroscopic confirmation by other surveys is shown. The black histogram are the objects correctly identified by our classifier and the red are the failed ones. The counts on y-axis are shown in log scale for clarity.}\label{fig:faintspec}
\end{figure}

 When matched with 2dF-SDSS LRG and QSO Survey \citep{2009MNRAS.392...19C} final spectroscopic quasar catalogue covering an area of 191.9 deg$^2$ resulted in 30,261 objects. Of these only 7,293 objects have spectroscopic confirmation from 2dF and of those, 97.8 per cent objects were correctly predicted.  The  correctly identified quasars in this data  includes two gravitationally lensed quasars, SDSS J1216+3529 and SDSS J0832+0404, from \citet{2008AJ....135..520O} and 20 damped Lyman alpha quasars from \citet{2009MNRAS.392..998E}. A distribution of the correctly identified and failed faint quasars in our catalogue that have spectroscopic confirmation is shown in Fig.~\ref{fig:faintspec}.
\begin{table}
\caption{Unconfirmed predictions in the catalogue.\label{tab:cross2}}
\begin{tabular}{lccccccccc}
\hline
&  \multicolumn{3}{c}{DBNN Predictions} \\
\cline{2-4} \\
Cat. Code           & Quasar       &
Galaxy     & Star  &
 Ref\\
\hline
CNDWF &365 &4 &23 & 25 \\
ARC & 38 &0 &0 & 26 \\
XMM2iS &3176 &27 &287 & 27 \\
ROSAT-FSC &24 &0 &9 &28 \\
XMMCOSMOS &88 &0 &8 & 29\\
\hline
\end{tabular}
\\ (25) Brand et al.2006; (26) Aslan et al.2010; (27) Xmm-Newton Survey Science Centre, C. 2008, Vizier Online Data catalogue, 9040, 0; (28) Veron-Cetty et al.2004; (29) Cappelluti et al.2009;
\end{table}

\par A photometric (Richards+2009) catalogue of about 1.2 million quasars was constructed by \citet{2009ApJS..180...67R} using an 8-dimensional photometric classification scheme. They used a Bayesian based kernel density estimator  to identify quasar candidates from SDSS DR6 with a limiting magnitude of $i$=21.3 and expected completeness of  $\sim$ 70\% to type 1 quasars. 

\par
\begin{figure}
{\centering
\includegraphics[scale=0.45]{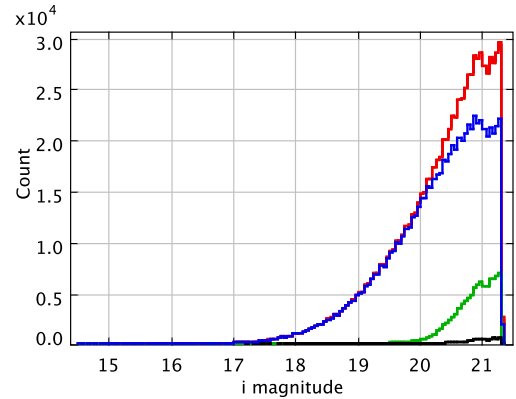}
\caption{The magnitude histogram of quasars from the photometric catalogue by Richards Gordon for objects with flag good $\geq$ 0 (red) is compared with the predictions of DBNN of the same as quasars (blue), stars (green) and galaxies (black).}
\label{fig:RPCmaghist}
}
\end{figure}

\begin{figure}
{\centering
\includegraphics[scale=0.45]{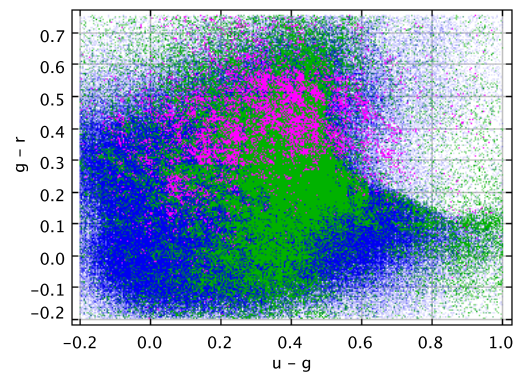}
\caption{A sample colour-colour plot of the distribution of objects in the photometric catalogue of Richards Gordon for objects with flag good $\geq$ 0 that are predicted by DBNN to be quasars (blue), stars (green) and galaxies (pink) is shown.}
\label{fig:RPCCC}
}
\end{figure}

 The Richards+2009 catalogue has 841,174 objects within the colour window used for our study. We used our trained classifier to determine the probable class of those objects and it labelled 734,803 (85 per cent) as quasars, 98,449 as stars and 7,922 as galaxies. A histogram of the distribution of the magnitudes of the matches shown in Fig.~\ref{fig:RPCmaghist} shows the good agreement between the catalogues at brighter magnitudes. The disagreement is mostly at fainter magnitudes. A colour-colour plot of the classified objects is shown in Fig. ~ \ref{fig:RPCCC}. To get some estimate of the quality of our classification, we compared these predictions with existing spectroscopic catalogues and give the completeness and contamination in tables \ref{tab:specmatchRPC} and \ref{tab:complt}. These tables show that our method gives better accuracy and lesser contamination in its prediction. Given that the data is already categorised as possible quasar candidates, a marginally higher contamination rate for stars and galaxies are understandable. This also explains why the number of stars in the table are much lesser than the number of quasars.

\begin{table}
\caption{A comparison of our predictions for objects with good $\geq$ 0 from Richards+2009 catalogue with objects that have spectroscopic confirmation from 2dF and other spectroscopic surveys. \label{tab:specmatchRPC}}
\begin{tabular}{|l|l|l|l|l|}
\hline
Confirmed & Quasars in & \multicolumn{3}{c}{DBNN classification as} \\
Spectral Class  & Richards+2009 & Quasar & Star & Galaxy \\
& Catalogue & & & \\
\hline \\
Quasar & 94858 & 94445 & 359 & 54 \\ 
Star & 1357 & 281 & 1074 & 2 \\
Galaxy & 323 & 91 & 22 & 210 \\
\hline \\
Accuracy  & 98.26 \% & 99.61 \% & 73.81 \% & 78.95 \%  \\
\hline
\end{tabular}
\end{table}

\begin{table}
\caption{Completeness and contamination details of predictions by DBNN as per Table \ref{tab:specmatchRPC}. \label{tab:complt}}
\begin{tabular}{|l|c|c|}
\hline
Object Type & Completeness \% & Contamination \%\\
\hline
Quasars & 99.56 & 0.39 \\
Stars & 79.15 & 26.19  \\
Galaxies & 65.02 & 21.05  \\
\hline
\end{tabular}
\end{table}
\subsection{Quasar Number Density}
\begin{figure}
{\centering
\includegraphics[scale=0.45]{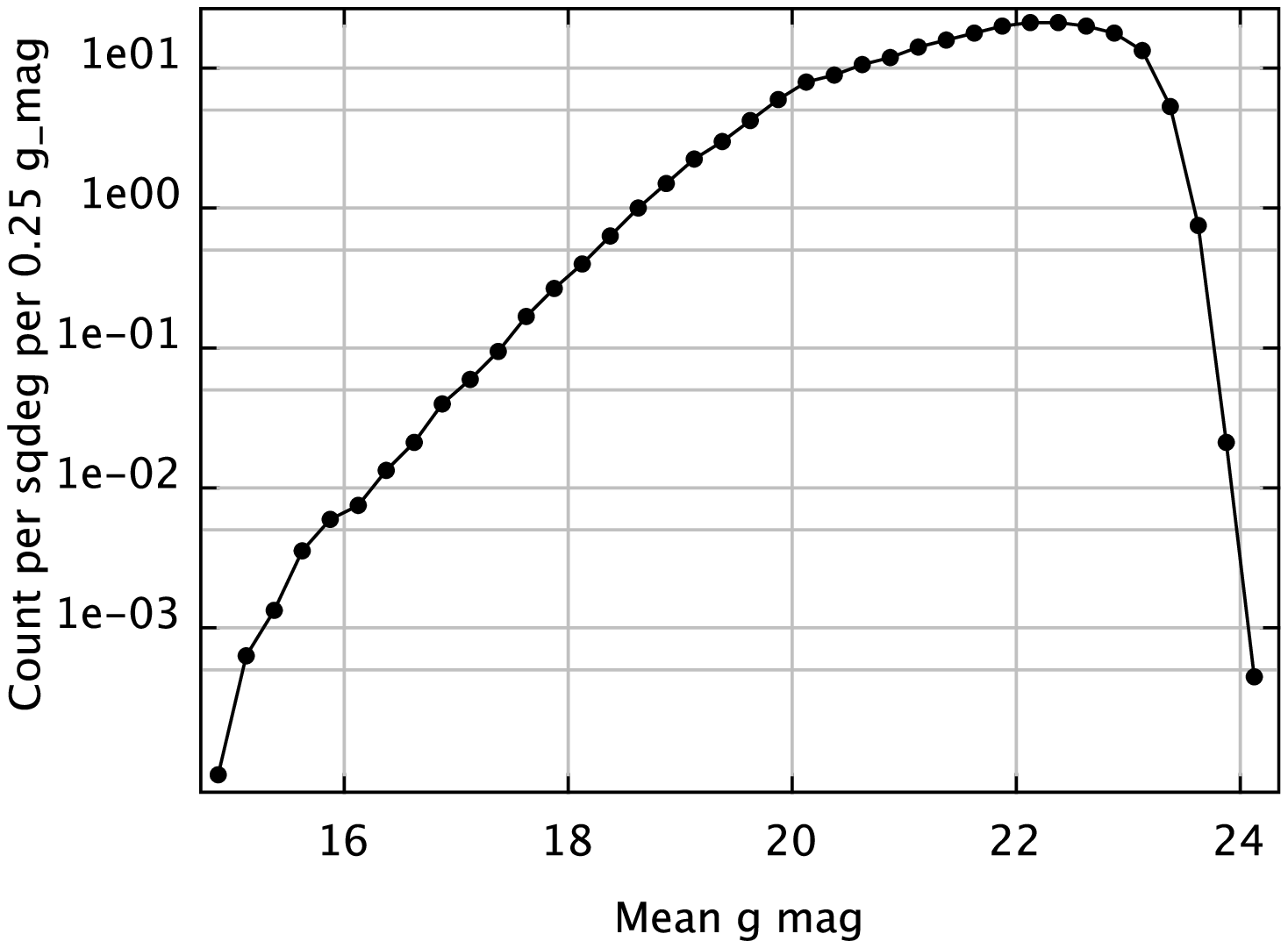}
\includegraphics[scale=0.45]{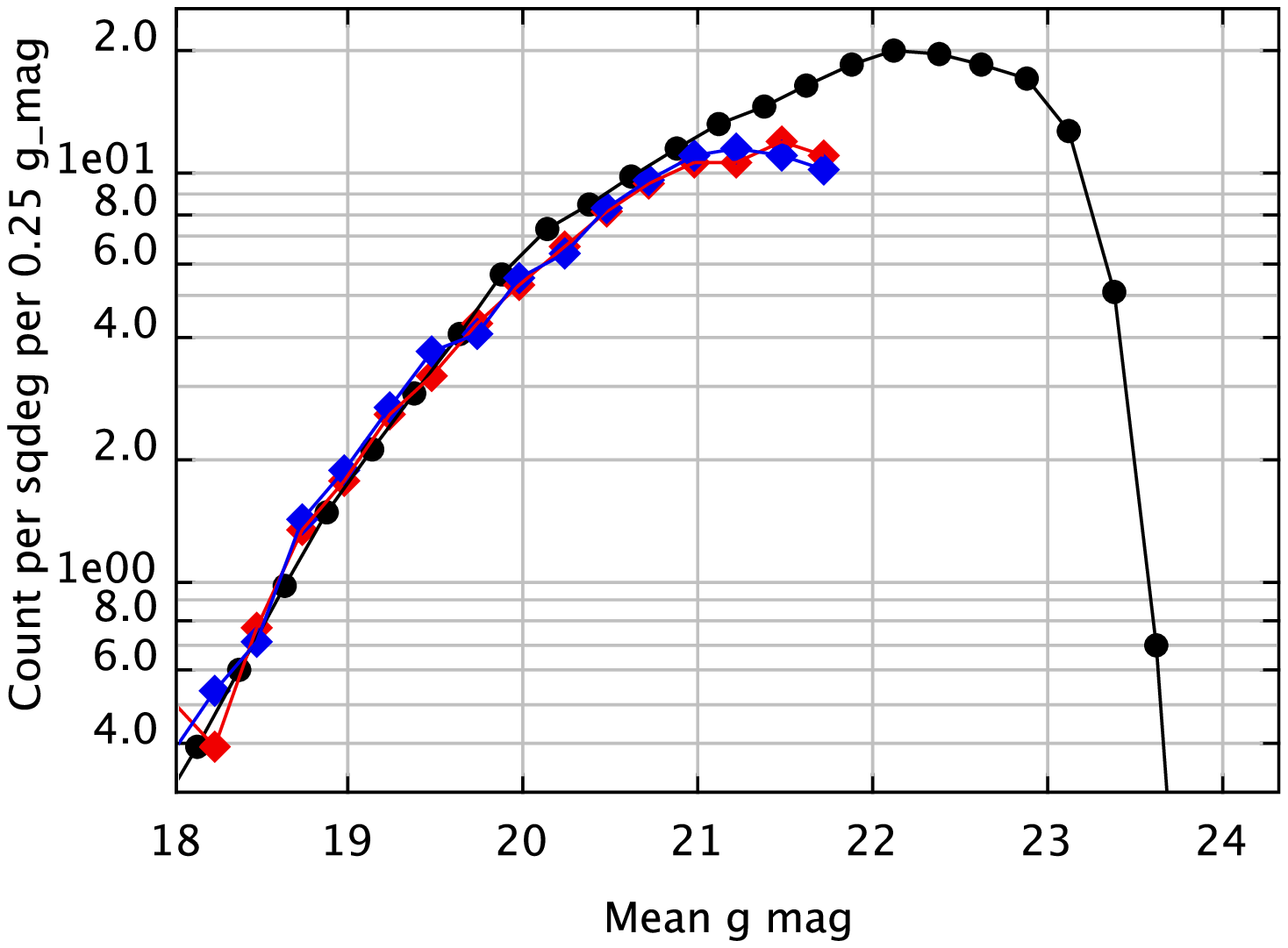}
\caption{The upper panel shows quasars number counts ($\Delta g =0.25$, $\sim 0.15 < z < \sim 2.6$) to the full magnitude range in our catalogue. In the lower panel, our predictions (black) are compared with observed quasar number count in the redshift range 0.4 to 2.1 from the 2SLAQ survey SGP (blue) and NGP (red) strips. Both show close agreement with our catalogue. The marginal deviations may be ignored considering the fact that our redshift window is $\sim 50$ per cent wider than ($z< 0.4, 2.1 < z < \sim 2.6$) the redshift coverage of the other two.}
\label{fig:QSPnumD}
}
\end{figure}

Our catalogue has about twice as many quasars as classified by Richards+2009. Does this mean that we are overestimating the quasar luminosity function? One way to evaluate the reliability of the catalogue is to compare the predicted quasar number density in our catalogue with what has been observed. Our catalogue of unresolved objects from $i \sim 14$ to 24 spans an area of 11,663 square degree with an expected coverage in redshift up to 2.6. Our study is restricted to unresolved objects to minimise the effect of the host galaxy on the colour estimates. \citet{2009MNRAS.399.1755C} have estimated the quasar luminosity function at $0.4 < z < 2.6$ based on 2dF-SDSS LRG and QSO (2SLAQ) survey having redshift range similar to our catalogue. In the upper panel of Fig. \ref{fig:QSPnumD}, the g-band quasar number counts for our complete catalogue and in the lower panel, the comparison with the g-band quasars number counts ($0.4 < z < 2.1$) from the 2SLAQ survey (after applying corrections for coverage, photometric and spectroscopic completeness) SGP (blue) and  NGP (red) strips respectively are shown. It may be noted that the counts agree at brighter magnitudes where as at fainter levels, our catalogue produces marginally larger counts because our redshift window is $\sim 50$ per cent wider, extending to $z \sim2.6$. We assume that our completeness for redshift $< 2.6$ is comparable to what has been depicted in Fig.\ref{fig:ug_gr} so that a simple comparison between the two is possible. The number density of quasars ($z \leqslant 2.6$) as per our catalogue is $\sim 116$ $deg^{-2}$ at limiting magnitudes $g\sim 22$ and falls to $\sim 54$ and $\sim 18$ quasars $deg^{-2}$ at g = 21 and 20 respectively. It is also noted that our number count, which most likely consists of quasars with $z < 2.6$, remains less than or equal to the redshift unbound number counts reported by \citet{1988ApJ...325...92K} at the respective magnitudes. Thus a factor in the excess of quasars in our catalogue as compared to Richard+2009, which is limited to $i \sim 21$, is to be understood as the contribution of objects from fainter magnitudes and as such is not contradicting the earlier estimates of the quasar luminosity function.

\subsection{Contamination and Completeness}
Completeness is defined as the ratio  $N_{cor}/N_{obj}$ where $N_{cor}$ is the number of correctly predicted objects (say quasars) and $N_{obj}$ is the actual number of objects in that class. Likewise, contamination is defined as the ratio of incorrectly labelled objects to the total number of objects in a class. It may also be defined as $1-$ Accuracy.
\par An ideal situation would be one where the completeness is 100 per cent and the contamination is zero. However, this is not possible in reality. We have discussed the difficulties in the classification of some of the objects and a reasonable solution would be one where the sample is 'complete' beyond a certain threshold and the contamination is the minimum. The classifier assigns a confidence value to every prediction it makes, which may be meaningfully used for this purpose.
\par An important point to consider here is the scientific goal at the end of the classification process. If we want to get a high level of completeness, then we would want to keep a low cut-off value for the confidence that can have any value between {$1/N$} and 1, $N$ being the number of predefined classes to be isolated. However, if we want to target quasars for spectroscopy, we might want to obtain quasar candidates that have a high chance to be a quasar. The selection of a suitable lower cut-off based on the confidence value allows one to do it. As it may be noted from the cumulative confidence plot (Fig.~\ref{fig:confhist}), a value of 55 per cent confidence could be regarded as a good cut-off for most purposes. However, this may also result in the loss of quasars at some specific patches of redshift where the confidence value drops because of the merger of colours from different classes. This is the inevitable trade-off between completeness and contamination.

\subsection{Future Plans}

Photometric estimation of redshift will significantly increase the usability of our catalogue by giving an additional dimension of their distributions. Secondly, the colour cuts that we used might have left out many interesting objects. Can we meaningfully classify objects from regions where there are fewer number of training samples? One option is to go to multi-band observations with capabilities to handle missing attributes. Many such objects are of specific  interest for target selection for astronomical observations such as variability and so forth. These and related issues are now being investigated.

\section{SUMMARY} \label{sum}
In this paper, we develop a machine learning algorithm based on Bayes theorem and train it on the colours of spectroscopically confirmed objects from SDSS to produce a catalogue of over 6 million unresolved photometric detections in the SDSS DR7, classifying them into stars, galaxies and quasars. These objects are from a small region of the SDSS colour space that has about 106,466 spectroscopically confirmed point sources and about 6 million photometric detections without spectroscopy, dominated by quasars and main sequence stars. We go to the limiting magnitudes of SDSS photometry and predict the class of the objects with a set of logically derived constraints. Our predictions are compared with other deep sky surveys in X-ray, optical, infra-red and radio and the results indicate that the method produces a reliable classification of faint objects using only the five SDSS magnitudes. The full catalogue and the data are available in electronic form. The high accuracy of matching and the ability to go to fainter levels of magnitude is expected to make our classifier a valuable addition to photometric classification and candidate identification for some of the upcoming deep sky surveys.
\par The catalogue is limited to the colour window used for this study and hence the completeness of the catalogue only refers to objects within this window. We have noted that many of the failures have occurred at specific patches of redshifts and are in agreement with literature. However we wish to note that the true nature of objects beyond $i\sim 21.3$ are unconfirmed and the artefacts in the data at fainter magnitudes could have resulted in inaccurate predictions by our classifier. The predictions give only an approximate estimate of the possible distributions at fainter magnitudes that may be confirmed and periodically improved when newer surveys spectroscopically confirm the nature of those objects.

\section*{Acknowledgements}
The authors wish to thank the anonymous reviewer for his valuable and detailed comments that greatly improved the quality of the manuscript. They also thank Scott M. Croom for providing the 2dF-SDSS LRG and QSO survey data for comparing with the catalogue quasar number counts in this work. This work was supported by ISRO Respond project grant No: ISRO/RES/2/339/2007-08. S.A and N.S.P wish to thank IUCAA for an extended stay and the use of the facilities for this work. The authors wish to express their sincere gratitude to S. G Ghosh and Asha Susan Jacob for proof reading the document. This study has extensively used SDSS data and the facilities they provide. Funding for the SDSS and SDSS-II was provided by the Alfred P. Sloan Foundation, the Participating Institutions, the National Aeronautics and Space Administration, the National Science Foundation, the U.S. Department of Energy, the Japanese Monbukagakusho, and the Max Planck Society. The SDSS web site is http://www.sdss.org/. We also wish to acknowledge the use of VOTools, mainly VOPlot and TOPCAT for this research.

\begin{verbatim}


The SQL query used to download the data from SDSS:

SELECT

s.objID as ObjID, 
s.ra as Ra, 
s.dec as Dec,
(u.psfMag_u - s.extinction_u) as psfMag_u,
(u.psfMag_g - s.extinction_g)as psfMag_g,
(u.psfMag_r - s.extinction_r) as psfMag_r,
(u.psfMag_i - s.extinction_i) as psfMag_i,
(u.psfMag_z - s.extinction_z) as psfMag_z,
s.type as type, 
s.z as RedShift, 
s.SpecClass as SpecClass 

FROM

UberCal u, 
SpecPhoto s into SpecUberMag 

WHERE

u.objID=s.objID and s.type=6 
AND
((u.psfMag_u  - s.extinction_u) -
(u.psfMag_g - s.extinction_g)) between -0.25 and 1.00 
AND 
(u.psfMag_g - s.extinction_g) -
(u.psfMag_r - s.extinction_r)) between -0.25 and 0.75
AND 
(u.psfMag_r - s.extinction_r) -
(u.psfMag_i - s.extinction_i)) between -0.30 and 0.50
AND 
(u.psfMag_i - s.extinction_i) -
(u.psfMag_z - s.extinction_z)) between -0.30 and 0.50
AND ((flags & 0x10000000) != 0)       
AND ((flags & 0x8100000c00a4) = 0)            
AND (((flags & 0x400000000000) = 0) or (psfmagerr_g <= 0.2))
AND (((flags & 0x100000000000) = 0) or (flags & 0x1000) = 0)
\end{verbatim}
\onecolumn
\begin{longtable}{cl}
\caption{Detailed description of the cross-matching results and the catalogues used \label{tab:discript}}
\\ \hline
\\
Cat. Code & Remarks\\
\hline
\\
2DF & The 2dF-SDSS LRG and QSO survey \citep{2009MNRAS.392...19C} is a spectroscopic quasar catalogue\\
&  which covers an area of 191.9 deg$^2$. There are 30,261 objects in 2dF catalogue that were \\
& overlapping with the colour space selected for our study. Of these, only 5510 2dF objects \\
& have spectroscopic confirmation. In that 4247 objects were predicted as quasars, \\
&  1113 objects as stars and remaining 238 objects were predicted as galaxies by our\\
& algorithm. Thus the overall accuracy is 98 per cent.\\

XBH**& This is the catalogue of 318 radio-quiet and X-ray emitting quasars (RQQ) studied by \\
& \citet{2008ApJS..176..355K}. Of 318 detections, 212 in the colour region investigated by us. \\
& All of them were correctly predicted by our classifier.\\

ASFS& The CRATES: All-Sky Survey of Flat-Spectrum Radio Sources \citep{2007ApJS..171...61H} has \\
& 14,467 sources of which 1131 overlap with the region of our analysis. These\\
& objects are characterized by a flat radio spectra with high variability in optical, significant\\
& polarization and bimodal synchrotron/Compton spectral energy distributions. Hence\\
& all of them are believed to be AGNs viewed 'pole-on'. The classifier correctly identified\\
& 1088 of them as quasars while it predicted 31 as stars and 12 as galaxies. \\

BATCS & \citet{2004AJ....127.2579Z} list the optical counterparts of 157 X-ray sources  selected using the multicolour \\
& CCD imaging observations made by the  Beijing-Arizona-Taiwan-Connecticut Sky Survey. Of these,\\
& 21 fall in the region of our catalogue and all are predicted as quasars while\\
& three of these objects were identified as star burst galaxies.\\

CGRBS** & CGRaBS is an all-sky gamma-ray Blazar candidates survey \citep{2008ApJS..175...97H} selected by \\
& their flat radio spectra. Of the 1625 target observations, 266 are in our catalogue. 265 of \\
& these were correctly classified as quasars while one got classified as galaxy. \\

DLyaQ**& \citet{2002PASA...19..455C} give a catalogue of 322 damped Lyman alpha absorbers. Of these 22 appear\\
& in our catalogue and 21 of them are correctly identified as quasars while one failed as a star.\\

F2QZ** & \citet{2005MNRAS.357.1267C} gives a sample of faint radio-loud quasars from FIRST. The sample\\
&  has 238 detections of which 190 appear in our catalogue. 186 of them are correctly\\
& identified as quasars and 3 failed as stars and one as a galaxy.\\

KFQS & \citet{2008MNRAS.386.1605M} have created a catalogue of 3154 $K$-band detections of possible quasar\\
&  candidates and their spectroscopic classifications. 159 of these objects appear in our\\
& catalogue while three of them have no spectral class. 144 of them were labelled as quasars. \\
& Two was predicted as galaxy and 13 as stars.\\

LQAC & \citet{2009AA...494..799S} have constructed a large quasar astrometric catalogue of 113666 quasars.\\
&  Of this, 61,788 overlap with our catalogue. It was found that our classifier correctly\\
&  detected 61,504 of them while predicting 267 as stars and 17 as galaxies.\\

LQRF & \citet{2009AA...505..385A} has constructed a large quasar reference frame of 100165 \\
& quasars observed in different surveys. Of these, 60,513 fall in the region of our \\
& catalogue objects. The classifier could correctly identify 60,280 of these objects\\
&  as quasars while 219 got labelled as stars and 14 as galaxies.\\

BZC** & Roma-BZCAT \citep{2009AA...495..691M} is a catalogue of 2837 blazars of which 255 had photometric \\
& detection by SDSS in the colour space of our catalogue. The classifier correctly detected 249 of \\
& them while got 4 incorrectly labelled as galaxies and 2 as stars.\\

PCS** &\citet{2004ApJS..150..165K} provide a catalogue of 220 spectroscopically confirmed AGNs using the\\
&  Faint Object Spectrograph on the Hubble space telescope. Of these, 55 $i$-band bright objects\\
&  appear in our catalogue. 53 of which are correctly identified as quasars while incorrectly labelled 2 as stars.\\

ROSA** &\citet{2006AJ....132.1475S} gives 1744 type 1 AGNs that have X ray observation in ROSAT PSPC. \\
& Of this, 1135 are present in our catalogue. All of these except 1 got correctly classified as quasars. \\

SQ13** & Quasars and Active Galactic Nuclei (13th Ed.) \citep{2010yCat.7258....0V} is a catalogue of\\
& 168,941 (all known prior to July 1st, 2009) AGNs. 65,673 of these objects have entries\\
&  in our catalogue of which 65,223 were correctly identified as quasars while 395 got labelled\\
&  as stars and 55 as galaxies.\\

SQR13 & Rejected Quasars and Active Galactic Nuclei (13th Ed.) \citep{2010yCat.7258....0V} \\
& has 178 entries that were previously considered as quasars and were rejected as mostly stars.\\
& Of these 28 objects have entry in our catalogue. Our algorithm incorrectly classified 7 of \\
& them as quasars while 21 were correctly labelled as stars.\\

DR7Q** & \citet{2010AJ....139.2360S} give the fifth edition of SDSS quasar catalogue consisting of 105,783 \\
& spectroscopically confirmed quasar candidates. Of these, the colour space covered by our catalogue \\
& contains 79,498 quasars. Our classifier correctly identified 79,140 quasars while it labelled \\
& 17 as galaxy and 341 as star.\\

SSSC & Skiff catalogue of Stellar Spectral Classifications \citep{2009yCat....102023S} is a compilation of 423055 \\
& stellar objects from literature. Of these, 1255 have entries in our catalogue. Our classifier \\
& labelled 82 objects as quasars of which 2 have been identified as quasars by SDSS DR7  \\
& quasar catalogue. Two objects got labelled as galaxy, while the remaining 1171 were \\
& correctly identified as stars.\\

SSA13 &\citet{2006ApJS..167..103F} prepared radio/optical catalogue of the SSA 13 Field that has 878\\
& radio sources of which only 6 have entries in our catalogue. All the objects except one were correctly \\
& picked by our classifier.\\

XMMSS & The second XMM-Newton Serendipitous Source catalogue \citep{2009AA...493..339W} has 3504 point  \\
& sources and 42 of these are in our catalogue. 37 of them were predicted as quasars  \\
& in that one is an emission line galaxy and 5 as stars. \\

SDSS/XMM & \citet{2009ApJS..183...17Y} gives the optical quasar candidates of 792 X-ray sources observed \\
& serendipitously in the X-ray with XMM Newton. 580 of these objects  appear in our catalogue \\
&  and all of them were correctly identified by the classifier.\\

RASS/2MASS & \citet{2009ApJS..184..138H} gives an associated catalogue of 18,806 X-ray sources in RASS/BSC \\
& that have counterpart with near-infrared sources from 2MASS/PSC. Of these, 6 objects appear \\
&  in our catalogue and all are labelled quasars by our classifier. \\

CAIXA & The catalogue of AGN in the XMM-Newton archive \citep{2009AA...495..421B} has 156 \\
& radio-quiet X-ray unobscured AGNs of which 16 appear in our catalogue. All of them \\
& got classified as quasars.\\

WDMB & \citet{2009AA...496..191H} gives 857 white dwarf - M binaries from SDSS DR6, of which 126 were present \\
& in our catalogue. 106 of them got labelled as stars while 20 were labelled as quasars.  \\
& One object predicted as quasar is a confirmed quasar from SDSS DR7 quasar catalogue. \\
& White dwarfs are known contaminants in photometric quasar catalogues\\
& explaining the relative low classification accuracy of 84 per cent in this case.\\

PMS & \citet{2004ApJS..152..103G} prepared a catalogue of proper motion of 390,476 stars from SDSS and\\
& USNO-B observations. 20,241 objects from it are present in our catalogue. Our classifier labelled \\
& 19,596 of the objects as stars, 639 as quasars and 6 as galaxies. Out of the 623 quasars predicted, \\
& 18 are confirmed quasars.\\

& \citet{2008AJ....135..520O} gave 4 gravitationally lensed quasars from SDSS Quasar Lens Search, \\
GLQ & which is a systematic survey of lensed quasars from SDSS  spectroscopic quasars. Of these, \\ 
&  two fall in the region of our catalogue and both are correctly predicted by the classifier. \\

CNDWF & The Chandra XBootes Survey optical counterpart catalogue \citep{2006ApJ...641..140B} has 5318 point \\
& sources and 392 of them are in our catalogue. The true class of the objects is not known. The\\
& classifier labelled 365 as quasars, 23 as stars and 4 as galaxies.\\

ARC & Astrometric positions of radio sources \citep{2010AA...510A..10A} give the positions of the  \\
& extragalactic radio detection of about 300 objects. Of this 38 are in our catalogue. All of them  \\
&  were classified as quasars.\\

XMM2iS & The XMM-Newton Second Incremental Source catalogue 
\\ & give a catalogue of 221,012 X-ray sources. Of these 3,490 overlap with our catalogue. Our classifier  \\
& identified 3,176 objects as quasars, 287 as stars and 27 as galaxies. \\

ROSAT-FSC & \citet{2004AA...414..487V} give optically selected bright AGN samples in ROSAT Faint Source \\
& catalogue. Of the 103 quasar candidate in this, 33 are present in our catalogue. 24 objects were correctly \\
& predicted by the classifier while 9 as stars. \\

XMMCOSMOS & The XMM - Newton wide-field survey \citep{2009AA...497..635C} in the COSMOS field gives \\
& 1887 point-like X-ray sources. 96 objects were present in our catalogue. Of these objects 88 \\
& were predicted as quasar while 8 as stars.\\ \\
\hline
\end{longtable}
\twocolumn

\end{document}